\documentclass[pra,aps,a4paper,showpacs,superscriptaddress,floatfix,twocolumn]{revtex4-1}
\usepackage{amsmath}
\usepackage{amsfonts}
\usepackage{amssymb}
\usepackage{graphicx}
\usepackage{dcolumn}
\usepackage{bm}
\begin{document}

\def\Zz{\mathbb{Z}}
\def\T{\hat{T}}

\def\beq{\begin{equation}}
\def\eeq{\end{equation}}
\def\nn{\nonumber}
\def\om{\omega}
\def\a{\alpha}
\def\b{\beta}
\def\g{\gamma}
\def\th{\theta}
\def\eps{\epsilon}
\def\veps{\varepsilon}
\def\l{\lambda}

\def\s{\sigma}
\def\D{\Delta}
\def\d{\delta}

\def\S{\hat{S}}
\def\ph{\phantom}
\def\ra{\rightarrow}

\title{Bound states in the one-dimensional two-particle Hubbard model with an impurity}

\author{J.~M.~Zhang}
\affiliation{Theoretical Physics III, Center for Electronic Correlations and Magnetism, University of Augsburg, 86135 Augsburg, Germany}

\author{Daniel Braak}
\affiliation{Experimental Physics VI, Center for Electronic Correlations and Magnetism, University of Augsburg, 86135 Augsburg, Germany}

\author{Marcus Kollar}
\affiliation{Theoretical Physics III, Center for Electronic Correlations and Magnetism, University of Augsburg, 86135 Augsburg, Germany}

\begin{abstract}
We investigate bound states in the one-dimensional two-particle Bose-Hubbard model with an attractive ($V> 0$) impurity potential. This is a one-dimensional, discrete analogy of the hydrogen negative ion H$^-$ problem. There are several different types of bound states in this system, each of which appears in a specific region. For given $V$, there exists a (positive) critical value $U_{c1}$ of $U$, below which the ground state is a bound state. Interestingly, close to the critical value ($U\lesssim U_{c1}$), the ground state can be described by the Chandrasekhar-type variational wave function, which was initially proposed for H$^-$. For $U>U_{c1}$, the ground state is no longer a bound state. However, there exists a second (larger) critical value $U_{c2}$ of $U$, above which a molecule-type bound state is established and stabilized by the repulsion. We have also tried to solve for the eigenstates of the model using the Bethe ansatz. 
The model possesses a global $\Zz_2$-symmetry (parity) which allows classification of all eigenstates into even and odd ones.
It is found that all states with odd-parity  have the Bethe form, but none of the states in the even-parity sector. 
This allows us to identify analytically two odd-parity bound states, which appear in the parameter regions  $-2V<U<-V$ and $-V<U<0$, respectively. Remarkably, the latter one can be \textit{embedded} in the continuum spectrum with appropriate parameters. Moreover, in part of these regions, there exists an even-parity bound state accompanying the corresponding odd-parity bound state with almost the same energy.
\end{abstract}

\pacs{03.65.Ge, 03.75.-b, 71.10.Fd}
\maketitle

\section{introduction}
The hydrogen atom serves as a paradigm in quantum physics. The exact solution \cite{schrodinger} of the two-body problem shows that one proton can bind one electron. However, a very interesting and highly nontrivial problem is the following: How many electrons can be added, if at all, to the hydrogen atom to form negatively charged ions? The answer, thanks to the endeavor of many people during more than half a century \cite{bethe, hylleraas,chan,hill,lieb}, is: \textit{at most one}, yielding  the fascinating object H$^-$ (for some very readable reviews, see \cite{rau01, bethebook}). The difficulty of this problem lies in the fact that the Coulomb repulsion between the electrons is of the same order of magnitude as the Coulomb attraction of the proton, which makes it a delicate job to balance the kinetic energy and the repulsion against the attraction to allow a bound state with two electrons below the continuum threshold. Actually, the repulsion between the electrons is so crucial that H$^-$ has only a single bound state \cite{hill}, i.e. the ground state, in which the two electrons are strongly correlated \cite{chan}. 

The problem  
can be generalized to any other atom: How many electrons can bind to a given nucleus of charge $+Z$? As a fundamental problem in both atomic and mathematical physics, it has attracted a lot of interest and many profound results have been established \cite{schope}.  

With the advent of the era of cold atoms, it becomes natural to pose similar questions for cold atoms in optical lattices. It is well-known that in a one-dimensional tight binding model, a defect potential $-V<0$ creates a localized mode around the defect \cite{feynman}. In the absence of atom-atom interaction, an arbitrary number of bosons can be trapped in this localized mode. However, as soon as an on-site atom-atom repulsion $U>0$ is turned on, at most a finite number of bosons can be trapped simultaneously, since the interaction energy scales with $N^2$ while the single particle energy scales with $N$, if $N$ bosons reside in the defect mode. The problem is then the following: What is the maximal value of $N$ for given values of $V$ and $U$? A different but related problem concerns the existence of bound states for fixed $N$ and varying $V$ and $U$. In particular one may ask whether the ground state is a bound state. 
Intuitively speaking, there must be a critical value of $U$ beyond which the ground state becomes delocalized by the repulsion.

These problems motivated us to study the bound states in a Bose-Hubbard model with an attractive impurity potential, for which some results were reported in Ref.~\onlinecite{prl2012}; in the limit of large $U$, $V$ a similar model was considered recently in Ref.~\onlinecite{santos}. Since numerical calculation is indispensable and since the dimension of the Hilbert space scales as $M^N$ with $M$ being the number of sites in a finite lattice, we shall only consider the second problem above and the case $N=2$. Therefore, we have a one-dimensional, discrete analogy of the H$^-$ problem. The attractive defect potential plays the role of the central Coulomb attraction while the atom-atom interaction corresponds to the Coulomb repulsion between the electrons. However, the possible experimental realizations with cold atoms allow more freedom in the choice of $V$ and $U$, which are freely adjustable. Especially, both of them can be either positive or negative. 

It turns out that this simple model embodies rich physics. For example, unlike the electron-electron repulsion in H$^-$, here the repulsion between the atoms plays a dual role. On the one hand, it may destroy a bound ground state, on the other hand, it allows formation of some other bound state. Specifically, for a given attractive potential ($V>0$), there exists a critical value $U_{c1}$ of $U$, above which the bound ground state is destroyed by the atom-atom repulsion. 
But for
$U \ge U_{c2} > U_{c1}$, a molecule-type bound state appears. This is a repulsion-aided bound state --- the two atoms are bound into a molecule by the repulsion and are trapped as a whole by the impurity potential. 
A noteworthy feature of this bound state is that it can be identified as a \textit{bound state at threshold} \cite{quantum,zeroeig,volcano,mattisrmp,hoffmann}, 
which is normalizable (thus of a finite size) even at the continuum threshold.

A most unexpected object realized in this model is a ``bound state in the continuum'' (BIC) \cite{wigner}. It is a normalizable state yet its energy 
is located within the continuum spectrum of the model. 
For $N=2$, the model seems simple enough to try a diagonalization via the Bethe ansatz \cite{betheansatz}, although the continuum version of the model is 
known to be nonintegrable \cite{mcguire}. 
It turns out that in our case a $\Zz_2$-symmetry (parity) divides the total Hilbert space into two sectors, one of which (odd under parity) allows diagonalization with the Bethe ansatz, while the other exhibits diffractive behavior. 
This enables us to construct the two odd-parity bound states in closed form, 
one of which can be a BIC. 

This paper is organized as follows. In Sec.~\ref{thebhm}, the model is defined and some of its basic properties are discussed. In Sec.~\ref{identification}, the numerical algorithm for identifying bound states in the system is introduced and the typical bound states found are displayed. Sec.~\ref{ground} and \ref{mole} treat the repulsive delocalization of the ground state and the repulsive localization of the molecule-type bound state, respectively. In particular, the values of $U_{c1,2}$ are determined both numerically and variationally, and the critical behavior of the bound states is investigated and compared. In Sec.~\ref{twoEX}, we discuss the Bethe ansatz solution of the model and the two analytically solvable odd-parity bound states. The accompanying bound states to the two analytically obtained bound states are the subject of  Sec.~\ref{sectionaccom}. We summarize our results in Sec.~\ref{sectionconclusion}. 

\section{the two-particle Bose-Hubbard model with a defect}\label{thebhm}

We consider two (spinless) interacting bosons (or equivalently, two spin-$1/2$ fermions in the spin singlet space) in a one-dimensional lattice with a site defect. The Hamiltonian of the system reads,
\begin{eqnarray}\label{h}
\hat{H} &=&\sum_{x=-\infty}^{+\infty}\left[ -(\hat{a}_x^\dagger \hat{a}_{x+1}+ \hat{a}_{x+1}^\dagger \hat{a}_x)+ \frac{U}{2} \hat{a}_x^\dagger \hat{a}_x^\dagger \hat{a}_x \hat{a}_x \right] \nonumber \\
& &\quad \quad \quad  -V\hat{a}_0^\dagger \hat{a}_0.
\end{eqnarray}
Here $\hat{a}_x$ ($\hat{a}_x^\dagger$) is the annihilation (creation) operator for a boson at site $x$. The Hamiltonian is just the conventional Bose-Hubbard model but with an impurity potential at site $0$. The value of the impurity potential is $-V$ while the on-site interaction between two bosons is $U$. Note that the hopping strength is set to unity. The Hamiltonian is invariant under the reflection about the defect $(\hat{a}_x,\hat{a}_x^\dagger)\rightarrow (\hat{a}_{-x},\hat{a}_{-x}^\dagger) $, which defines the parity of each state. 
Furthermore, the canonical transformation $(\hat{a}_x,\hat{a}_x^\dagger)\rightarrow (-)^x (\hat{a}_x,\hat{a}_x^\dagger)$ relates the eigenvalues and eigenstates of the Hamiltonian with parameters $(V,U)$ to those of the Hamiltonian with parameters $(-V,-U)$. Therefore, we shall consider in the following only the attractive impurity case $V>0$. The Hilbert space is spanned by Fock states which are defined as $|x_1,x_2\rangle \equiv \hat{a}_{x_1}^\dagger \hat{a}_{x_2}^\dagger |0\rangle $ if $x_1< x_2$, or $|x_1,x_2\rangle \equiv \frac{1}{\sqrt{2}} \left( \hat{a}_{x_1}^\dagger \right)^2|0\rangle$ if $x_1=x_2$.

The Hamiltonian (\ref{h}) is 
written in  second quantized form. For purposes below, it is also useful to work in the first quantization formalism. In this formalism, the wave function will be denoted as $f(x_1,x_2)$ with $ x_{1,2} $ being two independent variables. The Bose symmetry requires that $f(x_1,x_2)= f(x_2,x_1)$. The corresponding wave function in second quantization is of the form $\psi=\sum_{x_1\leq x_2} \psi(x_1,x_2) |x_1,x_2\rangle $, with
\begin{equation}
\psi(x_1,x_2)=  \left[\sqrt{2}-(\sqrt{2}-1)\delta_{x_1,x_2} \right]f(x_1,x_2).
\end{equation}
The Hamiltonian is defined by its action on the wave function
\begin{eqnarray}\label{h22}
\hat{H} f(x_1,x_2)=-\sum_{j=\pm 1} \left[f(x_1+j,x_2)+ f(x_1,x_2+j) \right ]\nonumber \\
\quad\quad \quad +\left[-V(\delta_{x_1,0}+ \delta_{x_2,0})+ U
\delta_{x_1,x_2} \right]f(x_1,x_2).\quad 
\end{eqnarray}
A state $f(x_1,x_2)$ is even/odd under parity if it satisfies
$f(x_1,x_2)=\pm f(-x_1,-x_2)$.

\begin{figure}[t]
\includegraphics[ width= 0.35\textwidth ]{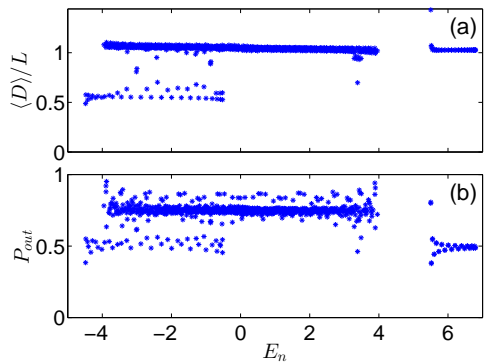}
\caption{(Color online) The three continuum bands revealed by using the two quantities of $D=|x_1|+|x_2|$ and $P_{out}$. The three horizontal lines are clearly visible in each panel. The eigenstates and eigenvalues $E_n$ are obtained by exact diagonalization on a $(2L+1)$-site ($L=20$) lattice and with open boundary conditions. The values of the parameters are $(V,U)=(1.5,5.5)$, and $R=L/2$.
\label{fig1}}
\end{figure}

\subsection{Three continuum bands}

We aim to identify the bound states, i.e., states in which the two particles are localized around the impurity. However, as a reference, it is necessary to identify the three continuum bands in this system first. 

The first band corresponds to the situation that the two bosons are neither captured by the impurity nor bound together by the interaction between them. For this band, the impurity and interaction cause phase shifts but do not contribute to the energy of the system, 
which is given
by the kinetic energy of the two bosons, and thus this band covers the interval $[-4,+4]$. 

For the second band, one boson is captured by the impurity yet the other is not. The energy of the first boson is $-\sqrt{V^2+4}$ \cite{feynman} while that of the other boson lies between $-2$ and $2$. 
This band covers therefore the interval $[-\sqrt{V^2+4}-2,-\sqrt{V^2+4}+2]$. 

The third band corresponds to a delocalized molecule state \cite{repulsive}: The two boson are bound together by the interaction between them and the composite moves as a whole on the lattice \cite{definition}. Again, the impurity causes phase shifts or local modifications on the wave function yet does not contribute to the energy. This band covers $[-\sqrt{U^2+16},U]$ if $U<0$ or $[U,\sqrt{U^2+16}]$ if $U>0$ \cite{repulsive}. 

The presence of the three bands can be demonstrated numerically by using some quantities which can reveal the distinct nature of the extended states in them. The first quantity is the sum of the distance of the two particles to the defect, $D=|x_1|+|x_2|$. Suppose we choose a lattice of $2L+1$ sites with $L$ sites on each side of the defect. For the first and third band, since the particles move through the lattice either independently or bound together, $\langle D \rangle $ should be $\sim L$. For the second band, since always one particle is localized around the defect, $\langle D \rangle $ is expected to be $\sim L/2$. Another quantity is the probability of finding at least one particle outside a ball with radius $R$ and centered at the defect, $P_{out}(f)=\sum_{\max\{|x_1|,|x_2|\}> R} |f(x_1,x_2)|^2$. If $R=L/2$, it is easy to see that $P_{out}$ should be $\sim 0.5$ for the second and third band, while for the first band it should be $ \sim 0.75$. These predictions are readily verified numerically. In Fig.~\ref{fig1}, we see how the three bands, although overlapping in energy, are separated by using $D$ and $P_{out}$. Moreover, their band edges coincide with the predicted values very well. 

Here some remarks are necessary. In \eqref{h} and \eqref{h22}, the Hamiltonian is defined on an infinite lattice, because bound states and extended states are only then well defined. However, in the numerical simulations, we can only deal with finite-sized lattices. We shall always choose a finite lattice with  $M=2L+1$ sites, open boundary conditions and the defect located in the middle, as in Fig.~\ref{fig1}. The Hamiltonian will be denoted as $\hat{H}_M$ and reads
\begin{eqnarray}\label{h33}
\hat{H}_M &=&-\sum_{x=-L}^{L-1} (\hat{a}_x^\dagger \hat{a}_{x+1}+ \hat{a}_{x+1}^\dagger \hat{a}_x)+ \frac{U}{2} \sum_{x=-L}^{L} \hat{a}_x^\dagger \hat{a}_x^\dagger \hat{a}_x \hat{a}_x \nonumber \\
& &\quad \quad \quad  -V\hat{a}_0^\dagger \hat{a}_0. 
\end{eqnarray}
Note that the defect site is set in the middle to preserve the reflection symmetry of the model. 
Thus, all the eigenstates have definite parity. In particular, it is easy to prove that the ground state 
[written as $|g\rangle $ or $f_g(x_1,x_2)$] is even and positive everywhere up to a global phase.

\section{Bound state identification}\label{identification}

The quantities  $D$ and $P_{out}$ can not only be used to identify the three different bands, but also to discern the bound states from the extended states \cite{ipr}. The latter are the overwhelming majority. The point is that there is an essential difference between bound states and extended states in their response to the change of the boundaries. An extended state would be very sensitive to the size of the lattice or the conditions (periodic or anti-periodic or in-between) imposed at the ends of the lattice: Its energy should change continuously as the boundary conditions change continuously and the typical amplitude of the wave function should decrease polynomially as the lattice size increases. On the other hand, the energy of a bound state should be independent of the lattice size or the boundary conditions, at least if the lattice size is much larger than the characteristic size of the wave function, since its wave function decays rapidly away from the center. 
In the case of extended states, the value of $\langle D \rangle $ would increase linearly proportional to $M $ (as argued in the previous section), and the value of $P_{out}$ (with a fixed $R$, however large it is) would converge to unity, as $M$ increases to infinity. But for a bound state, the value of $\langle D \rangle $ would converge to a finite value and the value of $P_{out}$ would converge to a value which is always smaller than unity and decrease with growing $R$. 

\begin{figure}[t b]
\includegraphics[ width= 0.35\textwidth ]{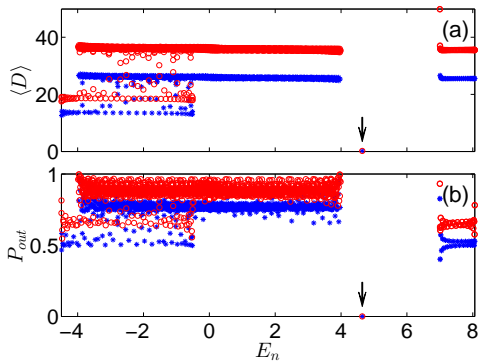}
\caption{(Color online) A bound state (indicated by the arrow in each panel, where a blue $\ast$ and a red $\circ$ coincide) found by using the lattice expansion algorithm. In each panel, the blue $\ast$ markers correspond to the smaller lattice with $51$ sites, while the red $\circ$ markers to the larger lattice with $71$ sites. The values of the parameters are $(V,U)=(1.5,7)$, and $R=12$.
\label{fig2}}
\end{figure}

\begin{figure*}[t b]
\begin{minipage}[b]{0.30 \textwidth}
\centering
\includegraphics[ width=\textwidth]{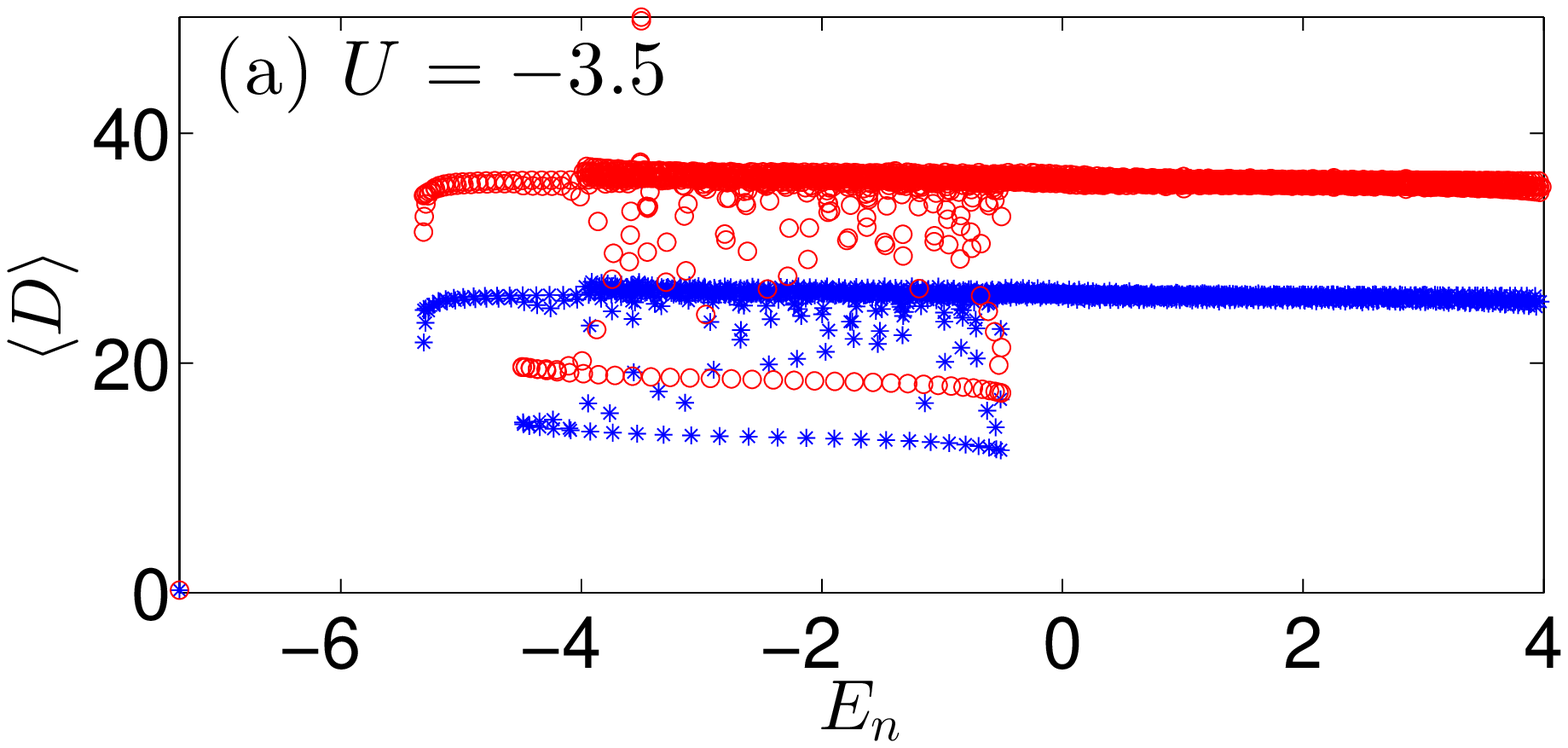}
\end{minipage}
\begin{minipage}[b]{0.30 \textwidth}
\centering
\includegraphics[ width=\textwidth]{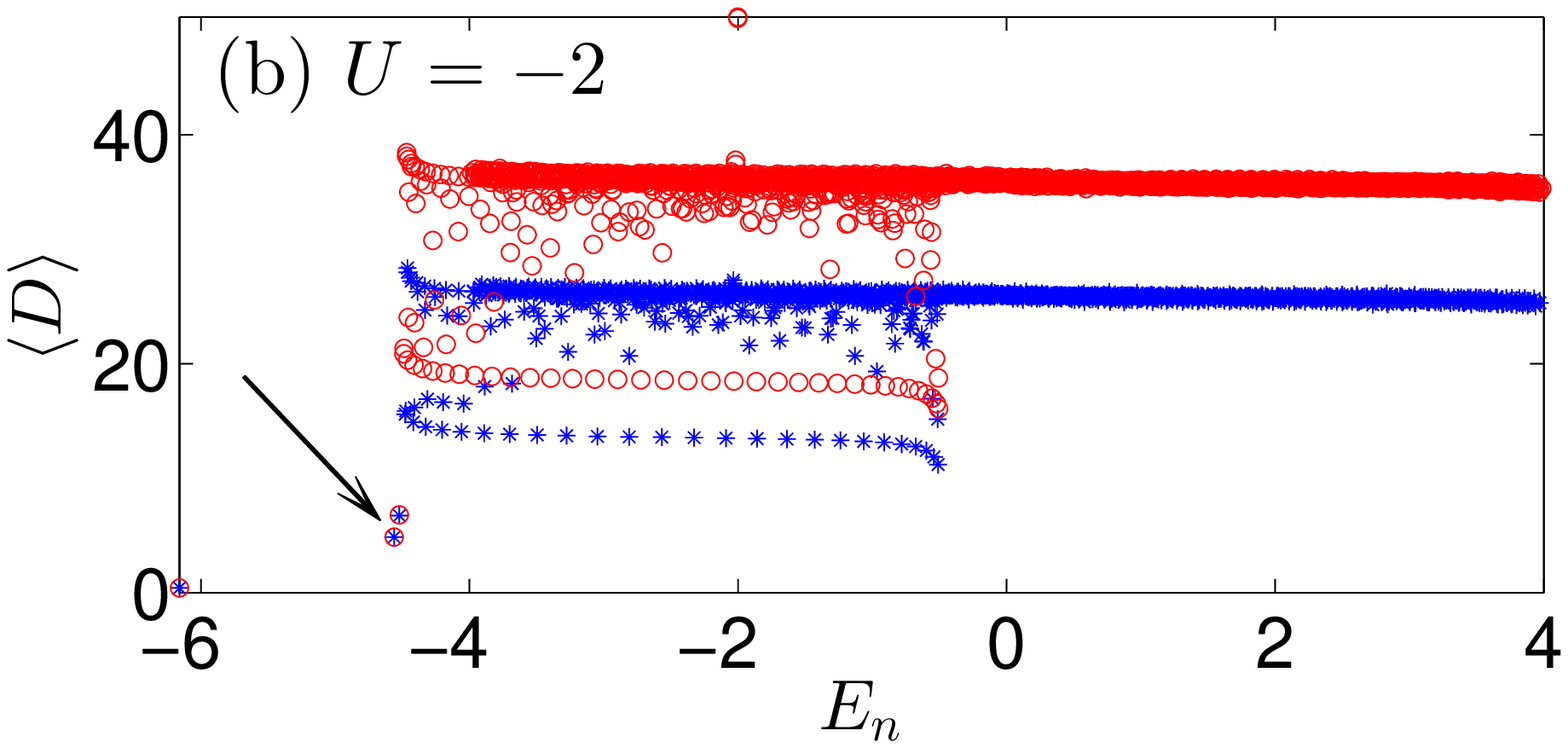}
\end{minipage}
\begin{minipage}[b]{0.30 \textwidth}
\centering
\includegraphics[ width=\textwidth]{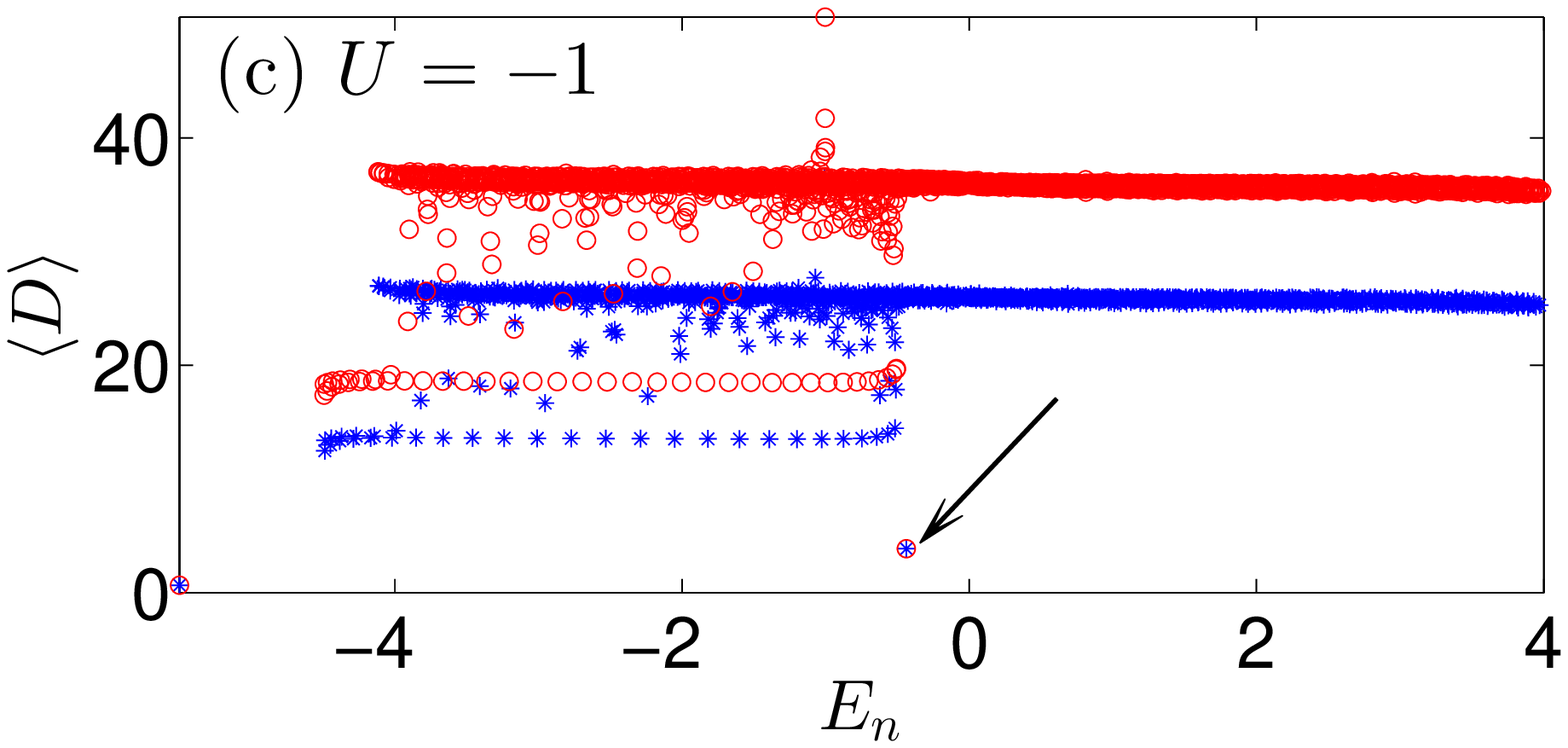}
\end{minipage}

\begin{minipage}[b]{0.30 \textwidth}
\centering
\includegraphics[ width=\textwidth]{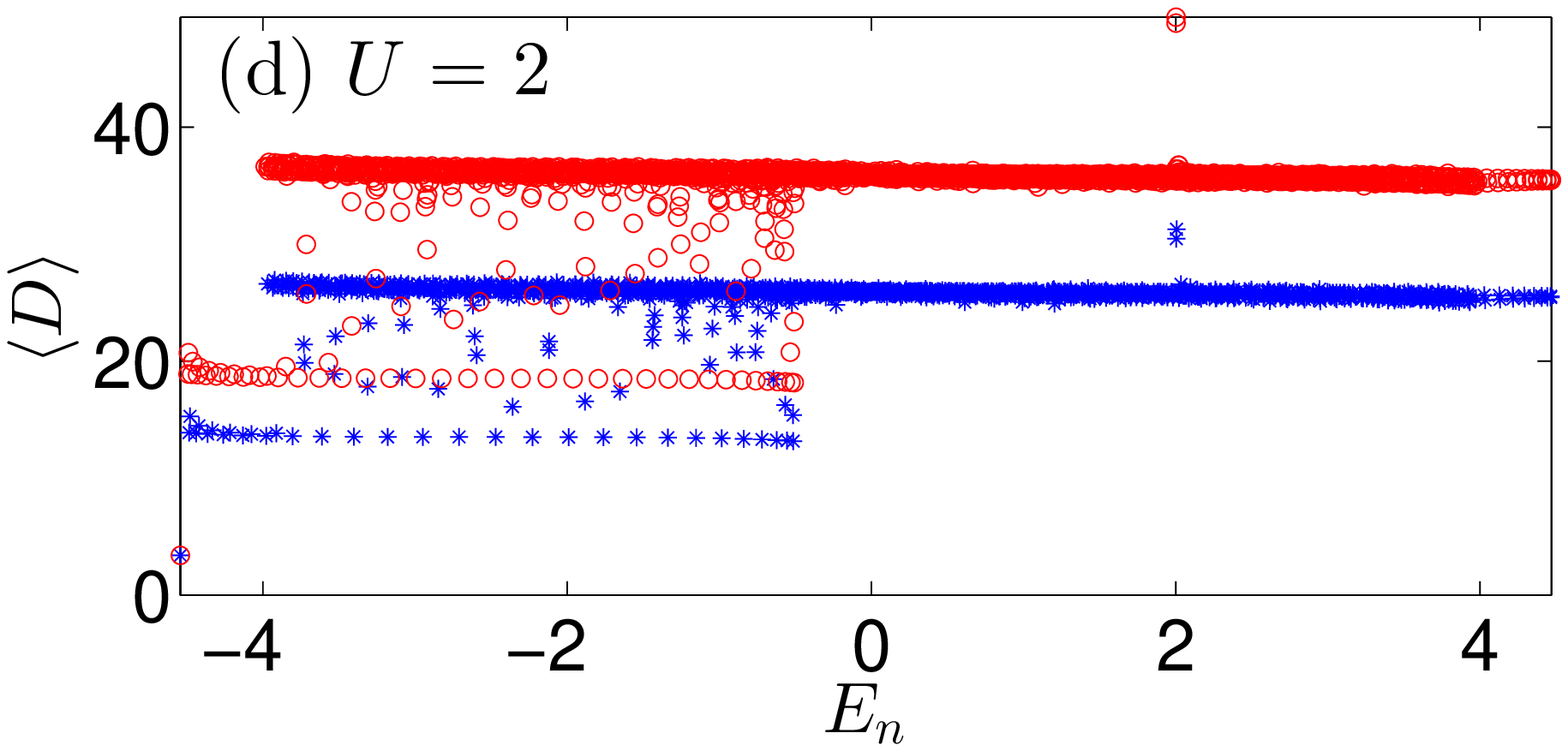}
\end{minipage}
\begin{minipage}[b]{0.30 \textwidth}
\centering
\includegraphics[ width=\textwidth]{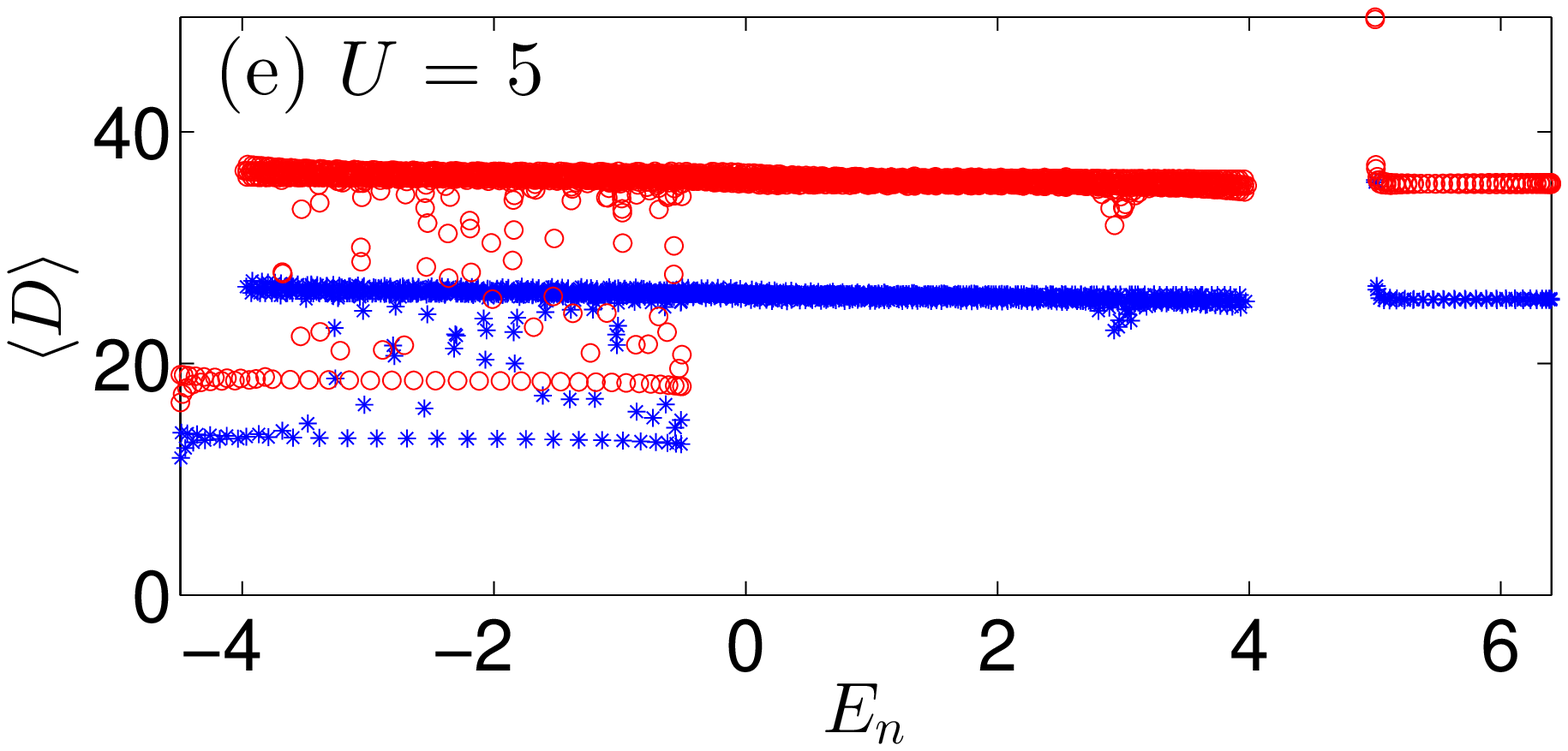}
\end{minipage}
\begin{minipage}[b]{0.30 \textwidth}
\centering
\includegraphics[ width=\textwidth]{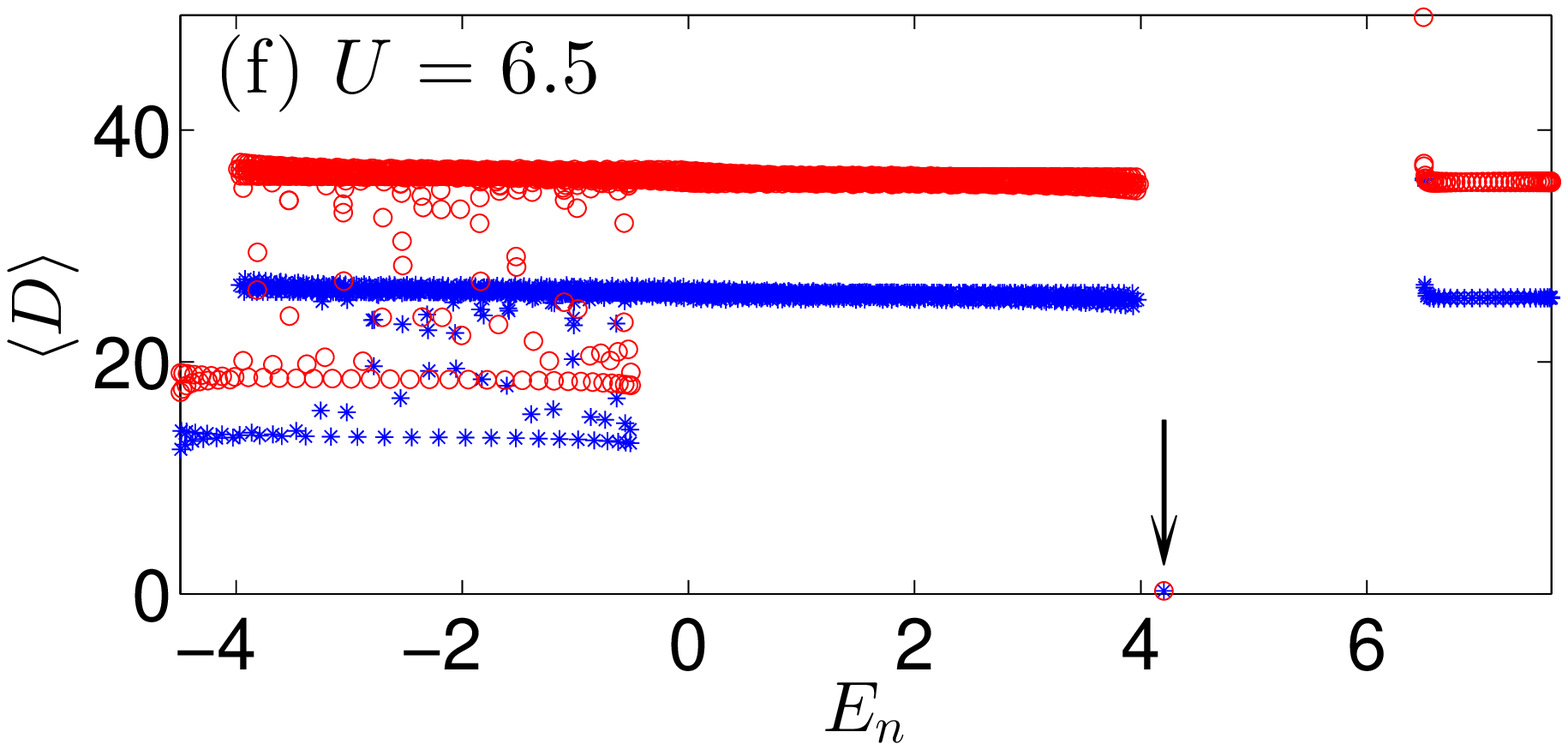}
\end{minipage}
\caption{(Color online) Typical bound states found by the filtering algorithm. From (a) to (f), the value of $U$ increases monotonically while the value of $V$ is fixed to $1.5 $. The sizes of the smaller (blue $\ast$) and larger (red $\circ$) lattice are $M_1=51$ and $M_2=71$, respectively. In each panel, a pair of blue $\ast$ and red $\circ$ which coincide with each other corresponds to a bound state. In (a), only the ground state is a bound state. In (b), besides the ground state, the first and second excited states (indicated by the arrow) are bound. In (c), besides the ground state, there is another bound state (indicated by the arrow) which is embedded in the continuum band. In (d), again only the ground state is bound. In (e), there is no bound state at all. In (f), a bound state (indicated by the arrow) appears in the gap between the first and third band. \label{fig3}}
\end{figure*}

Thus, to identify a bound state, we use the following algorithm: Begin with a  lattice of $M_1=2L_1+1$ sites with the defect site in the middle, and open boundary conditions. Solve all the eigenstates and eigenenergies of $\hat{H}_{M_1}$ numerically by exact diagonalization \cite{ejp}, and calculate the corresponding values of $D$ and $P_{out}$ for each eigenstate. Denote the $n$-th eigenstate as $\psi^{(M_1)}_n$, its eigenenergy as $E^{(M_1)}_n$, and the corresponding values of $D$ and $P_{out}$ as $D^{(M_1)}_n$ and $P^{(M_1)}_n$, respectively. Here $1\leq n \leq dim(M_1)$, with $dim(M_1)=M_1(2M_1+1)$ being the dimension of the Hilbert space on the $M_1$-sited lattice. The same procedure is then repeated on a (sufficiently) larger lattice of $M_2=2L_2+1$ sites, with $L_2>L_1$. A bound state would be signified by a pair of fixed values of $D$ and $P_{out}$, at some fixed eigenenergy. 
More precisely, there should exist two indexes $n_1$ and $n_2$ associated with each bound state, such that $(E^{(M_1)}_{n_1},D^{(M_1)}_{n_1},P^{(M_1)}_{n_1}) \simeq (E^{(M_2)}_{n_2},D^{(M_2)}_{n_2},P^{(M_2)}_{n_2})$, respectively.

As an illustration, we have carried out this algorithm for a specific pair of values of $(V,U)$ on a smaller lattice ($L_1=25$) and a larger lattice ($L_2=35$). The resulting values of $D$ and $P_{out}$ versus the eigenenergies are plotted in the upper and lower panels of Fig.~\ref{fig2}, respectively. There we see that as the lattice is enlarged, the markers in the three continuum bands move upwards significantly, in accord with our expectation. However, in each panel, in the gap between the first and third band and at the same energy, we have a blue marker coinciding with a red marker, which indicates a bound state. Actually, it is a repulsion-aided bound state, which will be discussed in detail in Sec.~\ref{mole} below. 

Finally, we would like to mention that the idea of distinguishing localized states from extended states by their different sensitivities to boundary conditions was employed before in the context of Anderson localization \cite{edwards}. Besides that, it was also proposed as a criterion of differentiating superfluid states from insulator states \cite{fisher}. A superfluid state would response to the twist of the boundary conditions by generating a finite current, while an insulator state would remain inert under this distortion.  

\subsection{A list of typical bound states}

With the effectiveness of the algorithm proven, we have explored the $(V,U)$ parameter space extensively. Some general behavior can be observed and is illustrated in Fig.~\ref{fig3} by taking a cut through the parameter space at fixed $V=1.5$ and varying $U$. We can see how the bound states change as $U$ is varied from large negative to large positive values. Only the quantity $D$ is displayed, but the same conclusions are obtained by using $P_{out}$.

(1) In Fig.~\ref{fig3}a, we see that for sufficiently large negative $U$, the ground state is a bound state and the only one. The fact that the ground state is localized is expected, since the interaction between the bosons strengthens the  binding  effect of the attractive potential.

(2) In Fig.~\ref{fig3}b, we see that as  $U$ is increased to $-2$, there are now two more bound states besides the ground state --- the first and second excited state (indicated by the arrow). Further investigation shows that the first excited state is always a bound state and is of odd-parity whenever $-2V<U<-V$. Actually, this bound state is analytically solvable by the Bethe ansatz. It will be discussed in detail in Sec.~\ref{twoEX}. As for the other bound state, detailed investigation shows that it appears only in a subset of the $-2V<U<-V$ region. In a certain sense, it is a companion of the analytically solvable one, always of even-parity and slightly higher in energy. It will be discussed in detail in Sec.~\ref{sectionaccom}.

(3) As $U$ is further increased to $-1$, as in Fig.~\ref{fig3}c, the two (excited) bound states disappear and are replaced by another bound state  (indicated by the arrow). This is no ordinary bound state since its energy falls in the $[-4,+4]$ continuum band. It is a ``bound state in the continuum'' (BIC). Further numerical and analytic investigation reveals that it appears whenever $-V<U<0$, and has  odd-parity and the Bethe form as the analytically solvable bound state in Fig.~\ref{fig3}b. They will be discussed together in Sec.~\ref{twoEX}. 

(4) As $U$ is increased to $2$, as in Fig.~\ref{fig3}d, the ground state is still a bound state, and the only one. However, its energy, which has been increasing since Fig.~\ref{fig3}a, is now very close to the lower edge of the second band. Here we should mention the simple fact that in the non-interacting case $U=0$, the ground state is the only bound state. It has the simple form $f_g(x_1,x_2) \propto \phi_d(x_1)\phi_d(x_2)$, with $\phi_d(x)\propto e^{-x/\xi}$ ($\xi=1/\ln[\frac{1}{2}(V+\sqrt{V^2+4})]$) being the single particle localized mode induced by the defect.

(5) As $U>0$ grows, a point is reached where the ground state is no longer bound. Actually, for  $U=5$, shown in Fig.~\ref{fig3}e, there is no bound state at all. We may conclude from Fig.~\ref{fig3}d to Fig.~\ref{fig3}e the existence of a critical value of $U$, denoted as $U_{c1}$, beyond which the ground state is turned into an extended one. 

(6) However, as $U$ becomes even larger, as in Fig.~\ref{fig3}f with $U=6.5$, a new bound state (indicated by the arrow) appears. It is located in the gap between the first and third band. In contrast to the ground state, this bound state is established and stabilized by the strong repulsion between the bosons.  
Obviously, there must be a second critical value $U_{c2}$ of $U$, above which this type of bound 
state becomes possible.  

\section{Stability of the ground state}\label{ground}

\begin{figure}[tb]
\includegraphics[ width= 0.45\textwidth ]{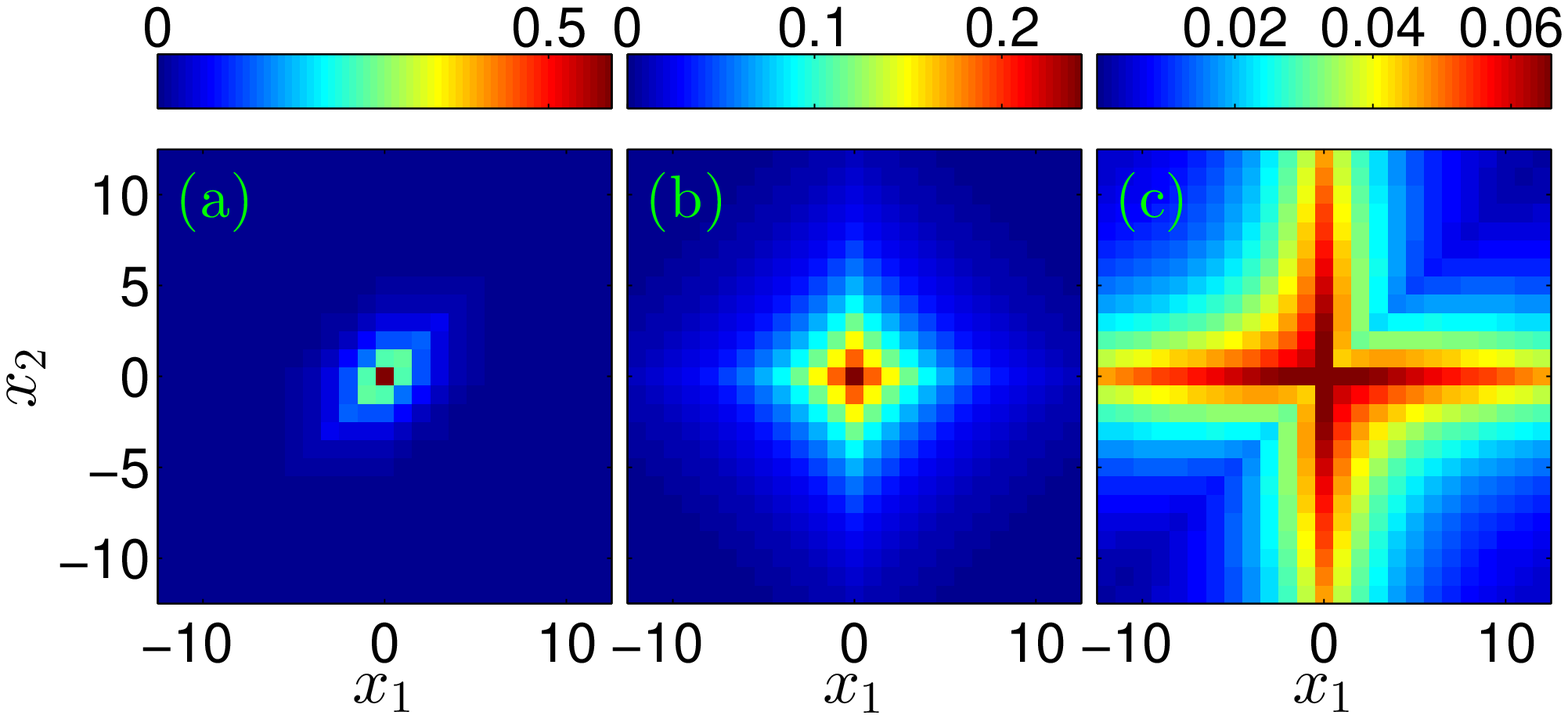}
\includegraphics[ width= 0.445\textwidth ]{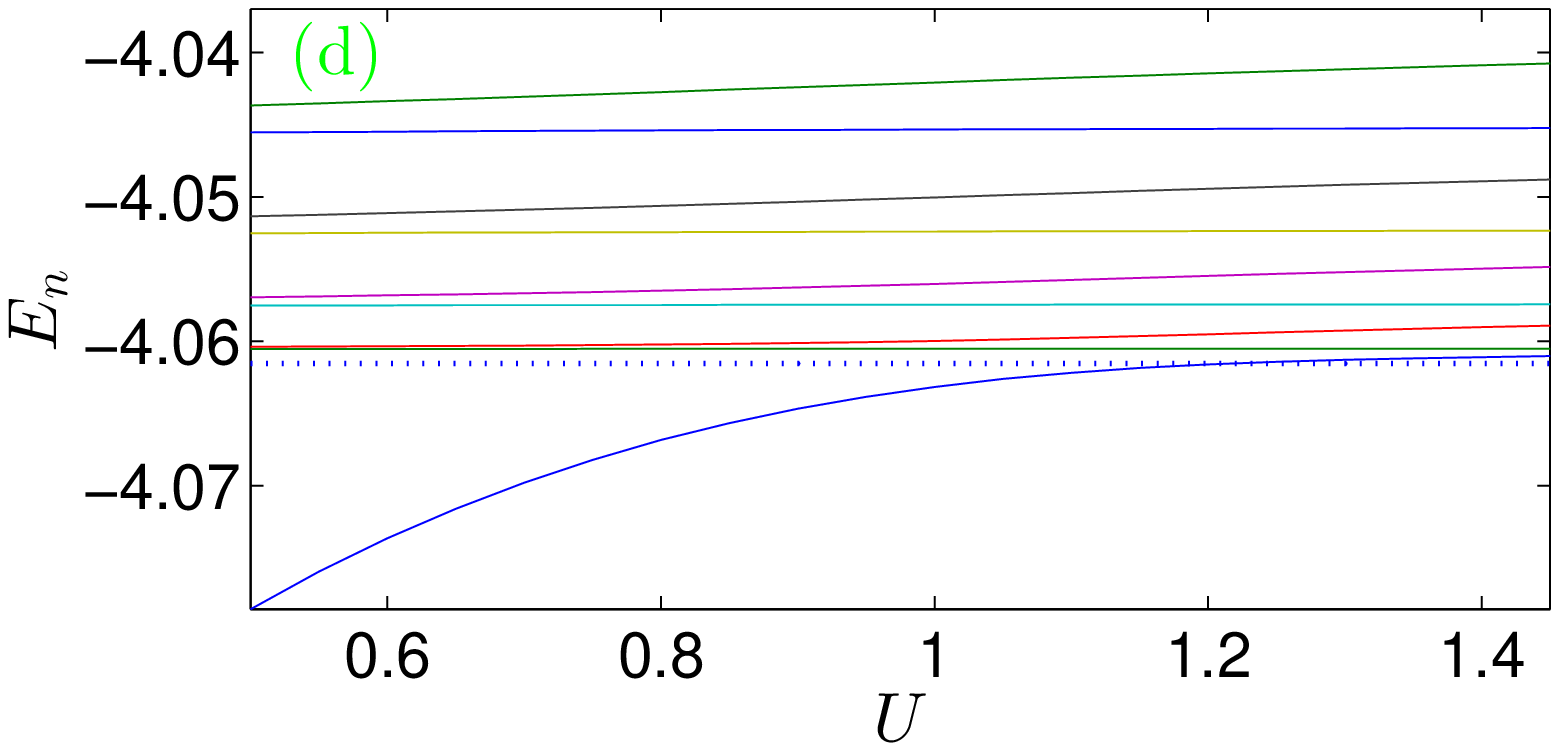}
\caption{(Color online) (a)-(c): Images of the ground state wave function $f_g(x_1,x_2)$ at three increasing values of $U$. Specifically, $U=-2$, $0$, and $1$ from left to right. (d): the bottom part of the spectral graph of the model as $U$ increases towards the first critical value $U_{c1}$. The value of $V$ is set to $0.5$. The lattice has 201 sites. In (d), the lowest blue line is the ground state level and the dotted horizontal line indicates the bottom threshold of the second continuum band $E_{edge2}= -\sqrt{V^2+4}-2 $. \label{gswave}}
\end{figure}

As expected and demonstrated in Fig.~\ref{fig3}, a sufficiently large $U$ can delocalize the ground state $|g\rangle $. In this section, we are interested in the critical value $U_{c1}$ as a function of $V$ and the critical behavior of the ground state. First, we note that for a positive value of $U$, 
the second band has the lowest lower band edge among all the three bands (see Fig.~\ref{fig1} as an example). Moreover, this upper limit for the ground state energy depends only on $V$ and it has the value  $E_{edge2}=-\sqrt{V^2+4}-2$. In terms of energy, it is then expected that as $U$ increases, the ground state energy increases according to the Hellmann-Feynman theorem, up to  the lower band edge of the second band and finally, at $U=U_{c1}$, it coalesces with it. This can already be seen in Fig.~\ref{fig3} and is most clear in Fig.~\ref{gswave}d, where the spectral graph of the model is shown.

More information is obtained by studying the evolution of the the ground state wave function $f_g(x_1,x_2)$ as $U$ increases. In Figs.~\ref{gswave}a-\ref{gswave}c, we have plotted the profile of $f_g(x_1,x_2)$ at three increasing values of $U$. In Fig.~\ref{gswave}a, we see that
for large negative $U$, the attractive defect potential is reinforced by the attractive interaction and the two bosons are sharply localized at the origin. In Fig.~\ref{gswave}b, for zero interaction $U=0$, $f_g$ is simply a product of the wave functions of the two bosons. The critical behavior of the ground state is exhibited in Fig.~\ref{gswave}c, where $U$ is close to $U_{c1}$. There, $f_g$ takes on a cross shape, which means that one boson is tightly bound by the defect, while the other is much more loosely bound. This indicates that the ground state wave function becomes delocalized continuously, since the wave functions in the second band have  the cross shape as well, except that they are infinitely extended along the axes. This behavior is quite reasonable physically: When one boson is localized around the defect, the other boson feels only a screened  potential, and thus is relatively loosely bound. 

Interestingly, a similar scenario occurs also for the hydrogen negative ion H$^-$. Due to the difficulty mentioned in the introduction, initially it took a great effort to prove H$^-$ as a bound system. However, Chandrasekhar had the insight that H$^-$ should have an ``in-out'' structure such that whenever one electron is close to the nucleus, the other is kept far away. He thus proposed a simple two-parameter trial wave function \cite{chan}, the essential part of which reads 
\begin{equation}
\chi(\vec{r}_1,\vec{r}_2)\propto e^{-\alpha r_1-\beta r_2}+ e^{-\alpha r_2-\beta r_1},
\end{equation}
where presumably $\alpha \gg \beta $. It turns out that the energy minimum is attained at $\alpha=1.03925$ and $\beta=0.28309$, and the corresponding energy is sufficient to prove binding for H$^-$. 

\begin{figure}[t b]
\includegraphics[ width= 0.4\textwidth ]{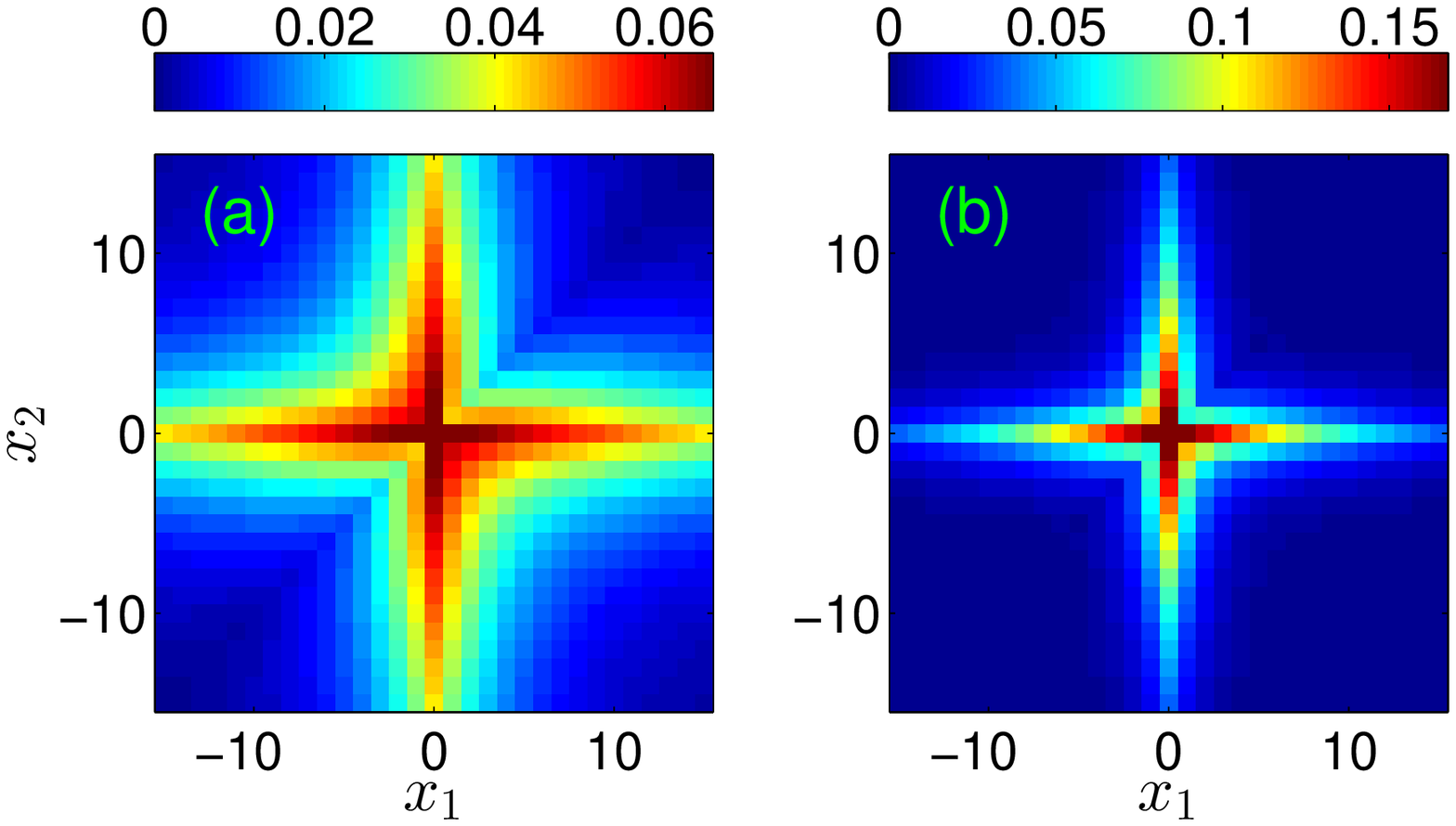}
\includegraphics[ width= 0.4\textwidth ]{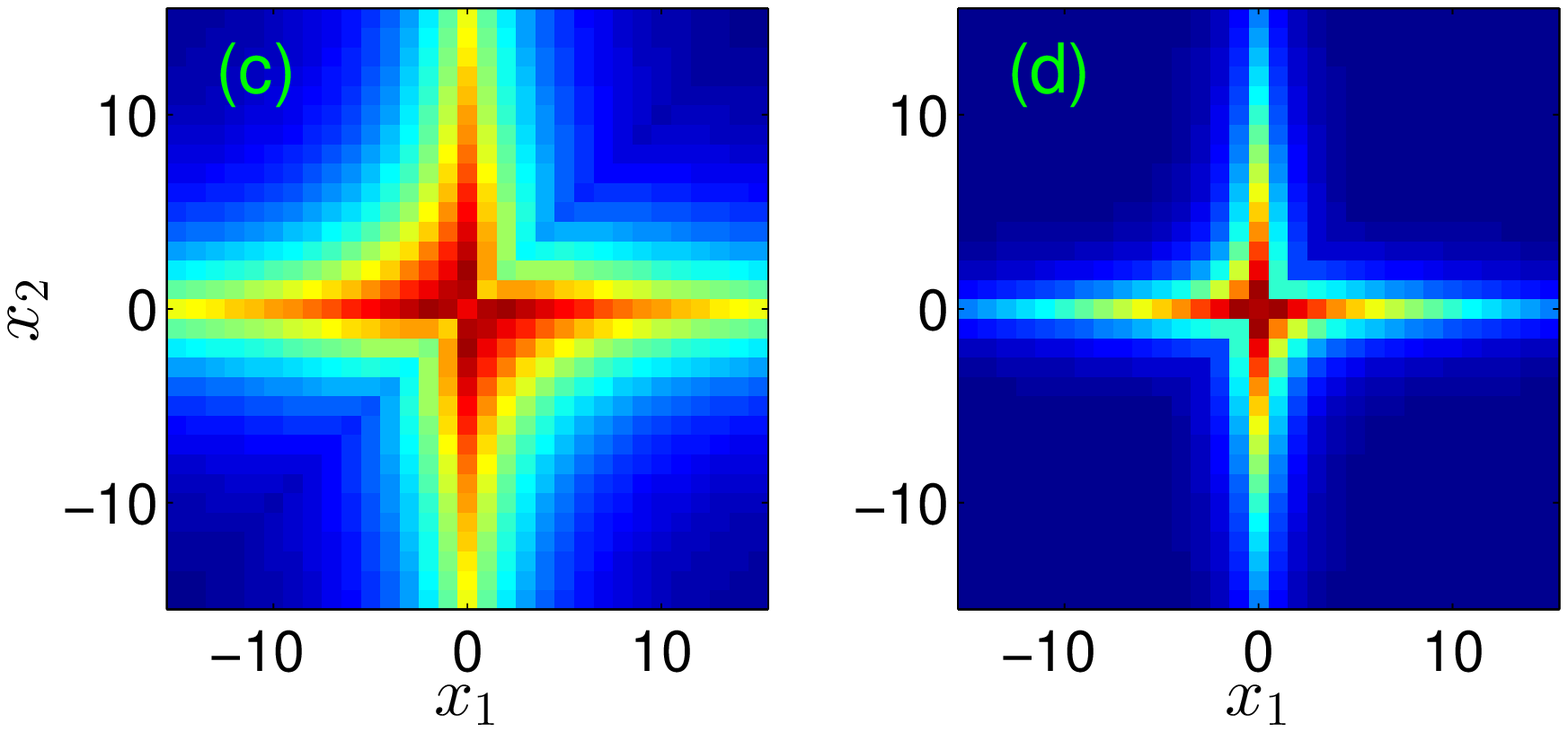}
\caption{(Color online) (a) and (b): images of the ground state wave functions obtained by exact diagonalization on a finite lattice with 301 sites. The parameters are (a) $(V,U)=(0.5,1)$ and (b) $(V,U)=(1.5,2.5)$. (c) and (d): images of the variational wave function minimizing the energy, with the same parameters as in (a) and (b), respectively. The inner product of the exact wave function and the variational one is 0.991 and 0.998, for (a)-(c) pair and (b)-(d) pair, respectively. \label{EDvsVAR}}
\end{figure}

Motivated by the observation of the critical behavior of the ground state and the similarity of the model with H$^-$, we have tried the following variational wave function for the ground state:
\begin{eqnarray}\label{variation}
f(x_1,x_2) &\propto & \left( e^{ -|x_1|/\lambda_1-|x_2|/\lambda_2}+ e^{ -|x_1|/\lambda_2-|x_2|/\lambda_1} \right) \nonumber \\
&& \times  \left( 1-s e^{-|x_1-x_2|/\lambda_3}  \right).
\end{eqnarray}
Here $\lambda_{1,2,3}$ are some characteristic lengths and $0\leq s\leq 1$ is an attenuation factor. We take the convention of $\lambda_1 \leq \lambda_2$. Here, besides the Chandrasekhar part (the first line), an additional factor (the second line) is introduced to suppress the wave function when the two bosons are close to each other. This is motivated by a fine structure of the wave function in Fig.~\ref{gswave}. That is, the value of the wave function at $(x,-x)$ is visibly larger than its value at $(x,x)$. It turns out that the ground state can be described by the variational form (\ref{variation}) quite well. In Fig.~\ref{EDvsVAR}, the ground states obtained by exact diagonalization are compared with those obtained within the variational scheme of (\ref{variation}). We see that they agree even in details. Quantitatively, the overlap between the exact and variational ground state exceeds 0.99 in both cases. Note that we did not treat $s$ as an independent parameter  in the variational calculation. Instead, it is determined as a function of $\lambda_3$, i.e., $s= \sqrt{(1-e^{-2/\lambda_3})/(1+e^{-2/\lambda_3})}$. However, the agreement is very good even without variation of $s$. 

\begin{figure}[t b]
\includegraphics[ width= 0.35\textwidth ]{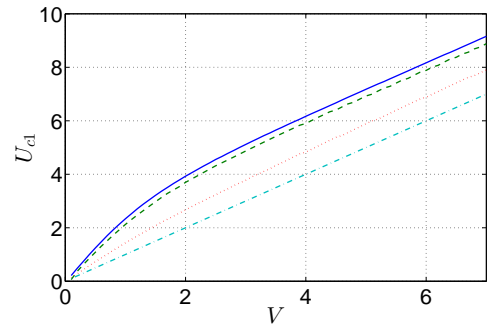}
\caption{(Color online) The bound-extended transition line of the ground state determined by various means. Blue solid line: exact-diagonalization on a 301-sited lattice; cyan dashed line: modified Chandrasekhar variational wave function (see Eq.~\eqref{variation}, $s\neq 0$); red dotted line: non-modified Chandrasekhar variational wave function ($s=0$); green dash-dotted line: $U_{c1}=V$ as in App.~\ref{variational}.\label{uc1}   }
\end{figure}

The critical value $U_{c1}$ can be determined efficiently by exact-diagonalization. We choose a $M$-sited lattice and using the Lanczos algorithm we can solve for the ground state energy $E_g(U,V;M)$ quickly. By the Hellmann-Feynman theorem, this quantity is a monotonically increasing function of $U$. Therefore, the critical value $U_{c1}$ is uniquely determined by the condition $E_g=E_{edge2}$. By using the Newton bisection method, it can be pinned down quickly. The $U_{c1}$ determined in this way depends of course on $M$. Actually, since a finite lattice can be understood as an infinite lattice with infinitely large potential beyond some radius, $E_g(U,V;M)$ is a monotonically decreasing function of $M$ by the Hellman-Feynman theorem. This means the $U_{c1}$ determined on a finite lattice is lower than the true value. However, for $M$ sufficiently large, the error is inappreciable. We have also tried to determine the transition line variationally:
If it is possible to find for a given $U$ a set of $\lambda_{1,2,3}$ and $s$ such that the energy of (\ref{variation}) is lower than $E_{edge2}$, then $U$ is a lower bound of $U_{c1}$. 
Both the exact diagonalization result (solid line) and the variational one (dashed line) are shown in Fig.~\ref{uc1}. There we see that the two approaches agree with each other very well in the full region of $V$ investigated. This again demonstrates that the wave function near the critical point can be well described by the variational form (\ref{variation}). 

The modified (or improved) Chandrasekhar wave function (\ref{variation}) is accurate but not very tractable analytically. Therefore, we have also tried some simpler variational schemes. The results are 
less accurate but capture well the qualitative behavior of $U_{c1}$ as a function of $V$. First, by a Hartree approximation as in App.~\ref{variational}, we can prove that $U_{c1} \geq V$. Second, by the non-modified Chandrasekhar (i.e., $s=0$) approximation as in App.~\ref{chandrasekhar}, we can show that $U_{c1}\geq 1.5 V$ as $V\rightarrow 0$ and $U_{c1} \geq V+1$ as $V\rightarrow +\infty$. These two results are also displayed in Fig.~\ref{uc1}. We see that they are not so accurate as (\ref{variation}), but still capture the behavior of $U_{c1}$ to  leading order in $V$ as $V\rightarrow +\infty$.
The fact that the non-modified Chandrasekhar wave function is less accurate than the modified one indicates that the suppression factor is quite essential near the critical point.

\section{Repulsion-aided bound state}\label{mole}

\begin{figure}[t b]
\includegraphics[ width= 0.45\textwidth ]{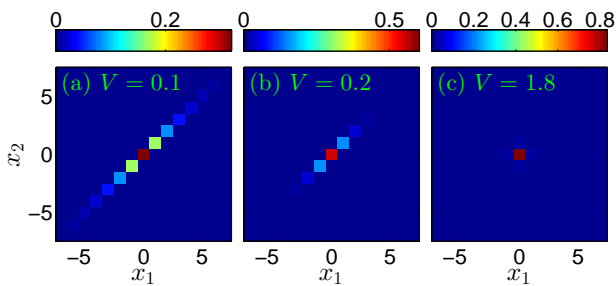}
\caption{(Color online) Images of the squared wave function $f^2_{mol}(x_1,x_2)$ of the molecule-type bound state with three different values of $V$ while the value of $U$ is fixed to $7$. The wave functions are obtained by exact diagonalization on a lattice of 201 sites. In (a) and (b), the wave function decays exponentially along the diagonal $x_1=x_2$. \label{molpic}   }
\end{figure}

In the preceeding section, we discussed the repulsion driven delocalization of the ground state. However, as revealed in Fig.~\ref{fig3}f, the repulsion can also facilitate the formation of a localized state. For a given $V$, there exists a critical $U$ (denoted as $U_{c2}$), beyond which a bound state appears in the gap between the first and third band. The mechanism for its formation relies on the repulsion between the two bosons which can bind them together to form a molecule (possible in a tight-binding model), which moves on the lattice as a whole with some effective hopping strength, and can thus be trapped by the defect potential. The picture is most clear in the large $U$ limit. In this limit, the double-occupied states $|x\rangle_m \equiv |x,x\rangle$ are degenerate and higher in energy than the non-double-occupied states by a gap of order $U$. Due to transitions through intermediate states like $|x,x+1\rangle $, two adjacent double-occupied states $|x\rangle_m $ and $|x+1\rangle_m$ are coupled by an effective hopping strength $2/U$, and each of them is up-shifted in energy by $4/U$. The effective Hamiltonian in the double-occupied subspace is then 
\begin{eqnarray}\label{effh}
\hat{H}_{eff}&=& \frac{2}{U}\sum_{x=-\infty}^{+\infty} (|x\rangle_{mm}\langle x+1|+|x+1\rangle_{mm}\langle x|)\nonumber \\
& & +  \left(U+\frac{4}{U} \right)\sum_{x=-\infty}^{+\infty} |x\rangle_{mm}\langle x| ,
\end{eqnarray}
which has the form of a standard tight-binding model. The dispersion relation reads $E_m(k)=U+\frac{4}{U}(1+\cos k)$, with a minimum at $k=\pi$. This molecule band is essentially the third band identified in Fig.~\ref{fig1}. Actually, its band range $[U,U+8/U]$ is approximately the exact one $[U,\sqrt{U^2+16}]$ in the limit of $U\rightarrow +\infty$. Now turn on the defect potential. If the potential is weak enough, we may still  confine 
the analysis to the double-occupied subspace. The defect is then equivalent to a term $-2V|0\rangle_{mm}\langle 0|$. As is well known \cite{feynman}, this leads to the formation of an even-parity bound state below the band minimum. Its energy is 
\begin{eqnarray}\label{emb1}
E_{mol} &=&  U+ \frac{4}{U}-\sqrt{4V^2 + \left( \frac{4}{U}\right)^2} \nonumber \\
& \simeq &   U -\frac{U}{2}V^2, \quad \text{if } V\ll \frac{2}{U}.
\end{eqnarray}
If $U>4$, $E_{mol}$ would be in the gap between the first and the third band (i.e. $4<E_{mol}< U$), for sufficiently small $V $. Thus, the bound state is expected to be stable. To verify the argument above, we show in  Fig.~\ref{molpic}a-b the squared wave functions of the bound state for two small values of $V$ (and a fixed value of $U$). The wave function is significantly different from zero only on the line  $x_1=x_2$ and decays exponentially away from $(0,0)$, which are two salient features in accord with the molecule picture. 

In the analysis above, the defect potential is assumed to be weak enough to allow  the restriction to the double-occupied subspace. If $V \gtrsim 1$, modifications are necessary. In this case, the ratio of the direct coupling ($= \sqrt{2}$) between the molecule state $|0\rangle_m$ and the disassociated states $|0,\pm 1\rangle$ to the gap between them ($U-V\simeq U$), is larger than the ratio of the effective coupling ($\simeq 2/U$) between the molecule states $|0\rangle_m$ and $|\pm 1 \rangle_m$ to the gap between them ($=2V$). Therefore, as a first approximation, we assume that the state $|0\rangle_m$ gets dressed by the state $\frac{1}{\sqrt{2}} (|0,+1\rangle+|-1,0\rangle)$ \cite{forbidden}. This is confirmed in Fig.~\ref{molpic}c, where $V$ is increased to 1.8. We see that the wave function is no longer extended in the $x_1=x_2$ direction, but takes on a cross shape---the values of the wave function $f$ at $(0,\pm 1)$ and $(\pm 1,0)$ are much larger than at $(1,1)$ or $(-1,-1)$. Diagonalizing the Hamiltonian in the two dimensional subspace of $\{ |0\rangle_m, \frac{1}{\sqrt{2}} (|0,+1\rangle+|-1,0\rangle) \}$, we get the energy of the bound state 
\begin{eqnarray}\label{emb2}
E_{mol} &=&\dfrac{1}{2}(U-3V)+ \sqrt{(U-V)^2/4+ 4} \nonumber \\
&\simeq & U-2V + \frac{4}{U-V}, \quad \text{if } U-V\gg 4.
\end{eqnarray}
Again, for sufficiently large $U$, $E_{mol}$ lies in the gap between the first and third band. 

\begin{figure}[tb]
\includegraphics[ width= 0.35\textwidth ]{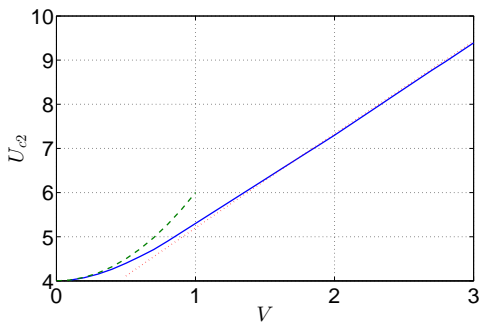}
\caption{(Color online) The second critical value of $U$ as a function of the defect potential strength $V $. The blue solid line is obtained numerically by using the lattice expansion algorithm, with a smaller lattice of 201 sites and a larger lattice of 301 sites. The green dashed (red dotted) line is the small- (large-) $V$ approximation as in Eq.~\eqref{uc2limits}. Note that $U_{c2}$ starts from 4.\label{fuc2}}
\end{figure}

By equating the value of $E_{mol}$ to 4, we obtain the critical value $U_{c2}$ 
for small and large $V$:
\begin{eqnarray}\label{uc2limits}
U_{c2}= \begin{cases}\quad 4+2V^2 , &V\rightarrow 0^+,\\ 4+2V-\dfrac{4}{V+4} , &V \rightarrow +\infty .\end{cases}
\end{eqnarray}
The value of $U_{c2}$ can also be determined numerically by using the lattice expansion algorithm. The result is shown in Fig.~\ref{fuc2} (the solid line). We see that as $V$ increases, the value of $U_{c2}$ interpolates smoothly between the small-$V$ and large-$V$ behavior in Eq.~\eqref{uc2limits}, with the crossover occurring around $V\simeq 1$. These results confirm the pictures above.

\begin{figure}[tb]
\includegraphics[ width= 0.36\textwidth ]{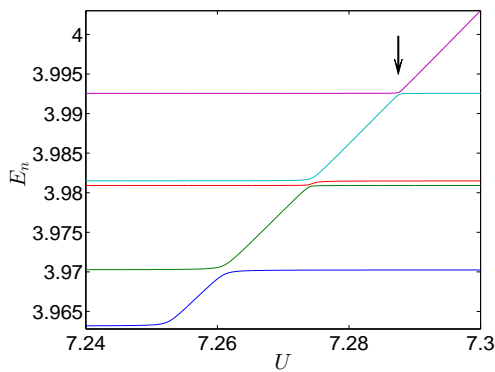}
\caption{(Color online) A portion of the spectrum (only even-parity levels are shown) of the model. The value of $V$ is set to 2. The lattice is of 101 sites with open boundary conditions. The highest one in the figure is the bound molecule one. The narrow anti-crossing is indicated by the arrow. Note the scales of the axes. \label{levelmol}   }
\end{figure}

\begin{figure}[tb]
\includegraphics[ width= 0.36\textwidth ]{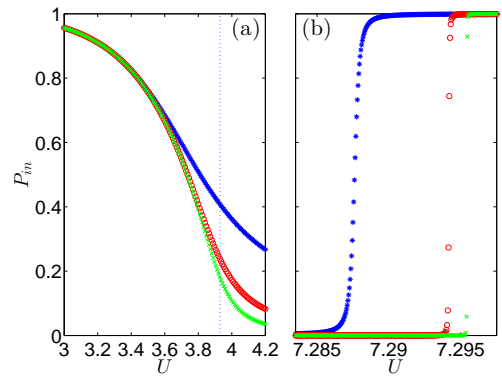}
\caption{(Color online) The probability of finding both bosons inside a radius of $R=10$, (a) in the ground state and (b) in the molecule-type bound state. In both panels, different markers indicate different lattice sizes (blue $\ast$: 101 sites; red $\circ$: 201 sites; green $\times$: 301 sites). The value of $V$ is set to $2$. The vertical dotted line in (a) indicates the critical value $U_{c1}$. Note the scale of the horizontal axis in (b). \label{conti}   }
\end{figure}

\subsection{A bound state at threshold}

We have shown that the repulsion can both destroy the ground state as a bound state and establish the molecule-type bound state. It is important to note that there is an essential difference between the two delocalization-localization transitions. The difference is, first of all, exhibited in the spectral graph. In Fig.~\ref{levelmol}, the molecule bound state energy at $U\sim U_{c2}$ is shown together with the few highest levels in the first band. In comparison with Fig.~\ref{gswave}d, where the ground state level merges into the second band smoothly without appreciable anti-crossing, here the molecule level merges into the first band by a very narrow anti-crossing. The different behavior of the levels implies that the first transition is a continuous one, while the second transition is essentially a discontinuous one. 
We have seen in Sec.~\ref{ground}, that during the first transition the wave function of the ground state becomes more and more extended along the $x_{1,2}$ axes
 as $U$ increases towards $U_{c1}$ from below. 
Eventually, at $U=U_{c1}$, the  characteristic length $\lambda_2$ diverges and the ground state turns into an extended state residing in the second band. The whole process is continuous. However, for the second transition, as $U$ decreases towards $U_{c2}$ from above, the wave function of the bound state is always of a finite (and small) size. Actually, the good agreement between  the two-level prediction [Eq.~\eqref{uc2limits}] and the exact value of $U_{c2}$ for large $V$  indicates that as long as the bound state is there, it can be well approximated using the two-dimensional subspace above. It happens in a very narrow interval of $U$ that the molecule bound state turns into an extended state, in which the two bosons are both delocalized and move virtually independently.

The two different pictures of the two transitions can be best demonstrated by using the quantity  $P_{in}=1-P_{out}$, i.e. the probability of finding both bosons within a radius $R$ from the defect. 
In Fig.~\ref{conti}a, $P_{in}$ is plotted versus $U$ for the ground state on three  lattices with different sizes. We see that $P_{in}$ is a continuous, monotonically decreasing function of $U$, and also a monotonically decreasing function of the lattice size. Arguably, on an infinite lattice, $P_{in}$ will drop continuously to 0 at $U_{c1}$. The behavior of $P_{in}$ is thus reminiscent of the order parameter in a second order phase transition. In contrast, in Fig.~\ref{conti}b, where the value of $P_{in}$ for the molecule-type bound state is plotted, $P_{in}$ transits from almost vanishing to almost unity in a very narrow interval. Moreover, the transition region gets smaller and smaller as the lattice size increases, which strongly indicates that the transition is \textit{discontinuous} on an infinite lattice This behavior is analogous to a first order phase transition. The discontinuity can be gleaned from Fig.~\ref{molpic} actually. There, for a given $U>4$, as $V$ increases from $0^+$ (transversing the phase diagram of Fig.~\ref{fuc2} from left to right), the wave function of the molecule-type bound state shrinks. This trend is incompatible with the configuration of the wave functions in the first band, where the two atoms are both delocalized and move virtually independently. Therefore, the transition between the two types of configurations cannot be continuous.

It is an intriguing fact that a bound state becomes delocalized abruptly. Though so far we have failed to prove this rigorously, we note that similar scenarios were noted previously in various single-particle \cite{quantum,zeroeig,volcano} or few-particle \cite{mattisrmp,hoffmann} systems, and especially, in the H$^-$ problem \cite{hoffmann}. It actually occurs in a simple and familiar single-particle model --- a three-dimensional finite depth square potential well \cite{quantum}. Due to the repulsive core created by the centrifugal barrier, a \textit{bound state with zero energy}  or a \textit{bound state at threshold} \cite{quantum,zeroeig,volcano,mattisrmp,hoffmann} is possible in this model for angular momentum $l\ge 1$. The remarkable property of this type of state is that it sits on the threshold to the continuum yet has \textit{finite} size. This means that if we adjust the parameter of the potential (e.g. the well depth) continuously across its critical value, we would observe a discontinuous behavior of $P_{in}$ similar to  Fig.~\ref{conti}b.

\section{two analytically solvable bound states}\label{twoEX}

In Fig.~\ref{fig3}c, we have a bound state \textit{inside} the $[-4,+4]$ continuum band. This is absolutely unexpected, since generally a bound state should lie outside of the continuum spectrum. The reason is best demonstrated by Mott's argument for the existence of sharp mobility edges \cite{mott}. In case of degeneracy between a localized state and an extended state, an infinitesimal perturbation would mix them and convert the localized state into an extended one. It means that although such exotic objects called ``bound state in the continuum (BIC)'' do exist, as pointed out by von Neumann and Wigner as early as in 1929 \cite{wigner}, they are expected to be unstable under perturbations. 

Nevertheless, it became recently possible to realize them experimentally in an electronic system \cite{albo}.
In our model, exploration of the $(V,U)$ space shows that the BIC is quite robust. It is always present 
whenever $-V<U<0$, though not necessarily inside the continuum. The robustness of the BIC indicates that it is protected by a hidden, perhaps even strong symmetry. This motivates the attempt to solve the model analytically using the Bethe ansatz. It turns out that this approach allows us to understand both the BIC in Fig.~\ref{fig3}c and one of the bound states in Fig.~\ref{fig3}b. 

\begin{figure}[t]
\includegraphics[width= 0.28\textwidth ]{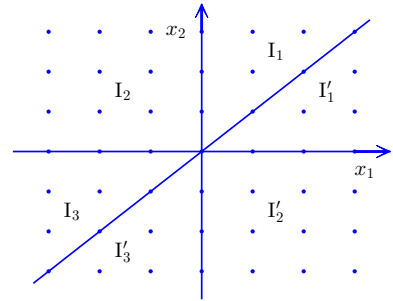}
\caption{(Color online) The $x_1$-$x_2$ lattice decomposed into six regions by the $x_1$-axis, $x_2$-axis, and $x_1=x_2$ line. Note that adjacent regions share the boundary between them. Especially, the origin belongs to all the six regions.
\label{lattice}}
\end{figure}

\subsection{The Bethe ansatz approach}\label{betheI}

We work with Eq.~\eqref{h22} and try to solve the eigenstates using the Bethe ansatz. An important observation is that the impurity potential and the interaction between the two particles are effective only on the three lines of $x_{1,2}=0$ and $x_1=x_2$. Away from these lines, we have free particles. This observation suggests to try an ansatz for the eigenstates of the Bethe form which is characterized by just two parameters, $k_1$ and $k_2$. 
Specifically, in region I$_1$ ($0\leq x_1 \leq x_2 $) in Fig.~\ref{lattice}, the wave function is postulated to read
\begin{eqnarray}\label{ansatz}
f(x_1,x_2)=\sum_{\sigma_0=0,1} \sum_{\sigma_1=0,1}\sum_{\sigma_2=0,1} A_{j(\sigma_0,\sigma_1,\sigma_2)}\quad\quad \quad \quad \quad \nonumber\\
\times \exp\left[i (-)^{\sigma_1} k_1 x_{1+\sigma_0}+i (-)^{\sigma_2} k_2 x_{2-\sigma_0}\right],\quad \,
\end{eqnarray}
where $ j(\sigma_0, \sigma_1,\sigma_2)\equiv 4\sigma_0+2 \sigma_1+\sigma_2+1 $. The wave function in regions I$_2$ and I$_3$ are defined similarly but with the $A$'s replaced by $B$'s and $C$'s, respectively. The value of the wave function in the other three regions is then determined by the condition $f(x_1,x_2)=f(x_2,x_1)$. In each region, we have eight different plane waves. The reason is that the interaction between the two particles can exchange their momenta, and the impurity potential can reverse the momentum of a particle. 

With the wave function in the form above, the eigenvalue equation $Hf=Ef$, with $E=-2\cos k_1 -2\cos k_2$, is satisfied away from the three lines. Now we just need to get it satisfied also on the three lines. This requires, besides the eigenvalue equation above, first of all, that $f$ be single valued on the interface between two neighboring regions. Specifically, on the boundary between I$_1$ and I$_2$, we have
\begin{subequations}\label{AB1}
\begin{eqnarray}
A_j+A_{j+2}=B_j+B_{j+2},\quad\quad\quad\quad\quad\quad \quad\\
 e^{ik_1} (A_j-B_j) + e^{-ik_1} (A_{j+2}-B_{j+2})\quad\quad\quad\quad\quad \nonumber\\
 \quad \quad =-V(B_j+B_{j+2}),\quad j=1,2, \quad 
\end{eqnarray}
\end{subequations}
and
\begin{subequations}\label{AB2}
\begin{eqnarray}
A_j+A_{j+1}=B_j+B_{j+1},\quad\quad\quad\quad\quad\quad \quad\\
 e^{ik_2} (A_j-B_j) + e^{-ik_2} (A_{j+1}-B_{j+1})\quad\quad\quad\quad\quad \nonumber\\
 \quad \quad =-V(B_j+B_{j+1}),\quad j=5,7. \quad 
\end{eqnarray}
\end{subequations}
On the boundary between I$_2$ and I$_3$, we have
\begin{subequations}\label{BC1}
\begin{eqnarray}
C_j+C_{j+1}=B_j+B_{j+1},\quad\quad\quad\quad\quad\quad \quad\\
 e^{ik_2} (B_j-C_j) + e^{-ik_2} (B_{j+1}-C_{j+1})\quad\quad\quad\quad\quad \nonumber\\
 \quad \quad =-V(B_j+B_{j+1}),\quad j=1,3, \quad 
\end{eqnarray}
\end{subequations}
and
\begin{subequations}\label{BC2}
\begin{eqnarray}
C_j+C_{j+2}=B_j+B_{j+2},\quad\quad\quad\quad\quad\quad \quad\\
 e^{ik_1} (B_j-C_j) + e^{-ik_1} (B_{j+2}-C_{j+2})\quad\quad\quad\quad\quad \nonumber\\
 \quad \quad =-V(B_j+B_{j+2}),\quad j=5,6. \quad 
\end{eqnarray}
\end{subequations}
On the boundary between I$_1$ and I$_1^\prime$, and the boundary between I$_3$ and I$_3^\prime$, we have
\begin{subequations}\label{CC}
\begin{eqnarray}
2i[(-)^m\sin k_1+(-)^n\sin k_2](A_{j+4} -A_j) \nonumber\\
= U(A_j+A_{j+4}),\\
2i[(-)^m\sin k_1+(-)^n\sin k_2](C_{j+4} -C_j) \nonumber\\
= U(C_j+C_{j+4}),
\end{eqnarray}
\end{subequations}
with $m,n=0,1$ and $j=2m+n+1+(-)^n$. Equations (\ref{AB1})-(\ref{CC}) constitute a set of 24 homogeneous linear equations of the 24 unknows $A$'s, $B$'s, and $C$'s. For this set of equations to have nontrivial solutions, the $24\times 24$ coefficient matrix, which depends on $k_1$, $k_2$, $V$, and $U$, should be singular. It is verified that the matrix is indeed always singular with a kernel of dimension 2. 

Here instead of dealing with the $24\times 24$ matrix, some simplification is possible. Note that the system is reflection invariant with respect to the impurity, which means if $f(x_1,x_2)$ satisfies the eigenvalue equation $Hf=Ef$, then so does $f(-x_1,-x_2)$. Therefore, we can classify the eigenstates into even and odd ones. 

For the even case, $f(x_1,x_2)=f(-x_1,-x_2)$, we need 
$B_i=B_{9-i}$ and $A_i=C_{9-i}$. Now, it is ready to verify that if we have $B_i=B_{9-i}$, then Eqs.~(\ref{AB1})-(\ref{BC2}) lead to $A_i=C_{9-i}$ automatically, and if Eq.~(\ref{CC}a) is satisfied, then (\ref{CC}b) is also satisfied. Solving the $A$'s in terms of the $B$'s from (\ref{AB1}) and (\ref{AB2}), and substituting the expressions into (\ref{CC}a), we get
\begin{widetext}
\begin{subequations}
\begin{eqnarray}
(U+2i(\sin k_1-\sin k_2))\left( B_1 - \frac{V}{2i \sin k_1}(B_1+B_3) \right) + (U-2i(\sin k_1-\sin k_2))\left( B_4 - \frac{V}{2i \sin k_2}(B_3+B_4) \right)=0,\quad \quad \\
(U-2i(\sin k_1-\sin k_2)) \left( B_4 +\frac{V}{2i \sin k_1}(B_2+B_4) \right)+ (U+2i(\sin k_1-\sin k_2)) \left(B_1 +\frac{V}{2i \sin k_2}(B_1+B_2) \right)=0,\quad \quad  \\
(U+2i(\sin k_1+\sin k_2)) \left( B_2 - \frac{V}{2i\sin k_1}(B_2+B_4) \right)+ (U-2i(\sin k_1+\sin k_2)) \left( B_3 +\frac{V}{2i\sin k_2}(B_3+B_4) \right)=0,\quad \quad  \\
(U-2i(\sin k_1+\sin k_2)) \left(B_3 +\frac{V}{2i\sin k_1}(B_1+B_3) \right)+ (U+2i(\sin k_1+\sin k_2)) \left( B_2 - \frac{V}{2i\sin k_2}(B_1+B_2) \right)=0.\quad \quad 
\end{eqnarray}
\end{subequations}
\end{widetext}
We thus get a $4\times 4$ matrix, which should be singular to admit non-trivial solutions of $B_1,B_2,B_3,B_4$. But the problem is that the matrix is always non-sigular as long as $UV\neq 0$, since its determinant is $-64U^2V^2$. Therefore, the even-parity eigenstates do not comply with the ansatz of \eqref{ansatz}.

For the odd case, $f(x_1,x_2)=-f(-x_1,-x_2)$, we need 
$B_i=-B_{9-i}$ and $A_i=-C_{9-i}$. Similarly, it is ready to verify that if we have $B_i=-B_{9-i}$, then Eqs.~(\ref{AB1})- (\ref{BC2}) lead to $A_i=-C_{9-i}$ automatically, and if Eq.~(\ref{CC}a) is satisfied, then (\ref{CC}b) is also satisfied. Solving the $A$'s in terms of the $B$'s from (\ref{AB1}) and (\ref{AB2}), and substituting the expressions into (\ref{CC}a), we get
\begin{widetext}
\begin{subequations}\label{oddb}
\begin{eqnarray}
(U+2i(\sin k_1-\sin k_2))\left( B_1 - \frac{V}{2i\sin k_1}(B_1+B_3) \right) - (U-2i(\sin k_1-\sin k_2))\left( B_4 -\frac{V}{2i \sin k_2}(B_3+B_4) \right)=0,\quad\quad \\
(U-2i(\sin k_1-\sin k_2)) \left( B_4 +\frac{V}{2i\sin k_1}(B_2+B_4) \right)- (U+2i(\sin k_1-\sin k_2)) \left(B_1 +\frac{V}{2i\sin k_2}(B_1+B_2) \right)=0,\quad\quad \\
(U+2i(\sin k_1+\sin k_2)) \left( B_2 -\frac{V}{2i\sin k_1}(B_2+B_4) \right)- (U-2i(\sin k_1+\sin k_2)) \left( B_3 +\frac{V}{2i\sin k_2}(B_3+B_4) \right)=0,\quad\quad \\
(U-2i(\sin k_1+\sin k_2)) \left(B_3 +\frac{V}{2i\sin k_1}(B_1+B_3) \right)- (U+2i(\sin k_1+\sin k_2)) \left( B_2 - \frac{V}{2i\sin k_2}(B_1+B_2) \right)=0.\quad\quad 
\end{eqnarray}
\end{subequations}
\end{widetext}
This time, the $4\times 4$ matrix is always singular and has a kernel of dimension 2. The value of 2 may be explained with   the time reversal symmetry of the Hamiltonian: If $f$ is an eigenstate, then so is $f^*$ with the same eigenvalue. This entails that for real $k_{1,2}$, if $(B_1,B_2,B_3,B_4)$ is a solution of (\ref{oddb}), then $(B_4^*,B_3^*,B_2^*,B_1^*)$ is also a solution of (\ref{oddb}).

Here some remarks are necessary. Suppose we allow both Bose and Fermi symmetry. Then Eq.~(\ref{h22}) itself is invariant under both the exchange of $x_1 \leftrightarrow x_2$ and the reflection of $x_{1,2}\rightarrow -x_{1,2}$. The two symmetries divide the Hilbert space into four sectors. For the anti-symmetric (fermionic) sectors, the interaction is ineffective and we have essentially free fermions in an impurity potential. The wave functions are in the Slater form and hence also in the Bethe form. Therefore, only in one of the four sectors, i.e. the symmetric (bosonic) sector with even parity, the wave functions are diffractive and not in the Bethe form. Note that the presence of diffraction in a related model defined on a continuous line was shown almost fifty years ago in a classic paper by McGuire \cite{mcguire}. The interesting point here is that the reflection symmetry decomposes the
bosonic Hilbert space into two subspaces, of which one shows no diffraction and can therefore be
considered integrable. This is confirmed by an analysis of the algebra of scattering matrices
and the associated Yang-Baxter relations, which is the subject of the next section.

\subsection{The integrable odd-parity subspace}\label{betheII}

We shall now formulate the problem in terms of scattering theory, usually employed in the analysis of
one-dimensional many-particle systems. This will explain the possibility to find eigenstates which have the Bethe form
although the bosonic model is nonintegrable even in the two-particle sector.
To this end we write the ansatz Eq.~(\ref{ansatz}) in all regions as follows: 
\begin{multline}
f(x_1,x_2)=\\
{\cal S}\sum_{R,\s_1,\s_2}\chi_R(x_1,x_2)A^R_{\s_1\s_2}(k_1,k_2)e^{i(\s_1k_1x_1+\s_2k_2x_2)}.
\label{BA}
\end{multline}
Here, $\chi_R(x_1,x_2)$ are characteristic functions of the six regions $R\in\{[i,j,k]\}$,
 \beq
 [i,j,k] \leftrightarrow x_i\le x_j\le x_k,\; i,j,k \in \{0,1,2\}, \;{\rm with }  \; x_0=0, \nonumber
 \eeq
while $\s_j\in\{1,-1\}$ denote a pseudospin degree of freedom accounting for the backscattering generated by the potential
 $-V$.
 Finally, the operator $\cal S$ symmetrizes the wavefunction with respect to $x_1$ and $x_2$.
The explicit use of $\cal S$ allows to reduce the number of amplitudes in each region from eight to four.
The eigenvalue equation $Hf=Ef$ entails then that the amplitudes 
 $A^R_{\s_1\s_2}(k_1,k_2)$ and
 $A^{R'}_{\s_1'\s_2'}(k_1,k_2)$ of adjacent regions $R,R'$ are related by the $S$-matrices,
 \beq
A^R_{\s_1\s_2}=\sum_{\s_1'\s_2'}\S^{\s_1'\s_2'}_{\s_1\s_2}(R,R')A^{R'}_{\s_1'\s_2'}.
\label{s-matrix}
\eeq
We have in general $\S(R,R')=\S(R',R)^{-1}$.
 The $S$-matrix corresponding to scattering between the two particles
 $\S([012],[021])$, taking particle 1 from the right to the left of particle 2, is diagonal in the pseudospin space
 due to conservation of total momentum,
 \beq
 S^{\s_1'\s_2'}_{\s_1\s_2}([012],[021]) = \d_{\s_1}^{\s_1'}\d_{\s_2}^{\s_2'}\ \frac{\sin\s_1k_1-\sin\s_2k_2 +iU/2}{\sin\s_1k_1
 -\sin\s_2k_2 -iU/2} \nonumber 
 \label{S12}
  \eeq
 and written as a $4\times4$ matrix, it reads,
 \begin{multline}
 \S_{12}(k_1,k_2)\equiv \S([012],[021])\\
=\S(120],[210])=\left(
 \begin{array}{llll}
 \a&0&0&0\\
 0&\b^{-1}&0&0\\
 0&0&\b&0\\
 0&0&0&\a^{-1}
 \end{array}
 \right)
 \label{S12-m}
 \end{multline}
 with
 \begin{subequations}
\begin{align}
\a&=\frac{\sin k_1-\sin k_2 +iU/2}{\sin k_1
 -\sin k_2 -iU/2},
\\
\b&=\frac{\sin k_1+\sin k_2 +iU/2}{\sin k_1
 +\sin k_2 -iU/2}.
 \label{s12-phases}
 \end{align}
\end{subequations}
 Note that $\S_{12}(k_1,k_2)^{-1}=\S_{12}(k_2,k_1)$ and  $\S_{12}$ is unitary for real $k_1,k_2$.

 The $S$-matrix describing the interaction between particle 1 and the potential $-V$ reads
 $\S_{10}(k_1)\equiv \S([102],[012])=\S([210],[201])$. It depends on $k_1$ but not on $k_2$ and has the form of
 \beq
\S_{10}(k_1)=
\left(
\begin{array}{llll}
1+\g_1 & \ph{1 - }\g_1&\ph{1+}0&\ph{1+}0\\
 \ph{1}-\g_1  & 1-\g_1&\ph{1+}0&\ph{1+}0\\ 
 \ph{1+}0&\ph{1+}0&1+\g_1 & \ph{1 - }\g_1\\
 \ph{1+}0&\ph{1+}0&\ph{1}-\g_1  & 1-\g_1  
 \end{array}
 \right)
 \label{S10-m}
 \eeq
with $\g_1=-iV/(2\sin k_1)$. Note that $\S_{10}$ 
is non-unitary (in contrast to $\S_{12}$) but preserves the current
 across the impurity:
 \beq
 \left|A^{[102]}_{1\s_2} \right|^2- \left|A^{[102]}_{-1\s_2}\right|^2 
 = 
  \left|A^{[012]}_{1\s_2}\right|^2- \left|A^{[012]}_{-1\s_2}\right|^2. 
 \label{current}
\eeq 
In an analogous way, we find for
 $\S_{20}(k_2)\equiv \S([201],[021])=\S([120],[102])$,
 \beq
 \S_{20}(k_2)=
\left(
\begin{array}{llll}
1+\g_2 & \ph{1 - }0&\ph{1+}\g_2&\ph{1+}0\\
 \ph{1+}0  & 1+\g_2&\ph{1+}0&\ph{1+}\g_2\\ 
 \ph{1}-\g_2&\ph{1+}0&1-\g_2 & \ph{1 - }0\\
 \ph{1+}0&\ph{1}-\g_2&\ph{1-}0  & 1-\g_2  
 \end{array}
 \right)
 \label{S20-m}
 \eeq
with $\g_2=-iV/(2\sin k_2)$.
Naturally, $[\S_{10},\S_{20}]=0$.

A necessary condition for the ansatz (\ref{BA}) to work is its self-consistency:
Starting from the amplitudes in one region, say $[021]$, we can compute the
amplitudes in all other regions using the $S$-matrices. The region $[120]$
can be reached from $[021]$ on two different paths, namely
$[021]\ra[012]\ra[102]\ra[120]$, which gives
\beq
A^{[120]}=\S_{20}\S_{10}\S_{12}A^{[021]}
\label{path1}
\eeq
or via $[021]\ra[201]\ra[210]\ra[120]$, corresponding to
\beq
A^{[120]}=\S_{12}\S_{10}\S_{20}A^{[021]}.
 \label{path2}
\eeq
 If an arbitrary eigenstate of (\ref{h22}) has the Bethe form (\ref{BA}), Eqs.~(\ref{path1}) and
 (\ref{path2}) would entail the matrix identity (Yang-Baxter equation),
 \beq
\S_{20}(k_2)\S_{10}(k_1)\S_{12}(k_1,k_2) =  
\S_{12}(k_1,k_2)\S_{10}(k_1)\S_{20}(k_2).\nonumber
 \label{ybe}
 \eeq
 Now, because $[S_{12},S_{10}S_{20}]\neq 0$, it is clear that the Bethe ansatz fails for
 general eigenstates. However, the matrix
 \beq
 \S_D=\S_{20}\S_{10}\S_{12}- \S_{12}\S_{10}\S_{20}
 \eeq
 has a two-dimensional kernel ${\cal K}=\langle A^{(1)},A^{(2)}\rangle$ for arbitrary $k_1,k_2,V,U$ and it follows
 \beq
 \tilde{A}^{(j)}\equiv \S_{20}\S_{10}\S_{12}A^{(j)}=\S_{12}\S_{10}\S_{20}A^{(j)},\quad j=1,2. 
\label{remamp}
 \eeq
If the amplitude vector in region $[021]$ is an element of $\cal K$, 
 we obtain along both paths the same amplitudes in region $[120]$ and the ansatz is consistent. 
This is also valid for the other 
 possible consistency relations
 because all regions are connected by a single loop and $\S(R,R')=\S(R',R)^{-1}$.  

To study the kernel of $\S_D$ further, we define
\begin{subequations}
 \begin{align}
    a &:= \gamma_1 \gamma_2 (1+\alpha) (1-\alpha^{-1}) \,,\\
    b &:= \gamma_1 \gamma_2 (1+\beta ) (1-\beta^{-1} ) \,,\\
    c &:= -(1+\gamma_2)\delta/\beta         \,,\\
    d &:= -(1-\gamma_2)\delta/\alpha        \,,\\
    e &:= -(1+\gamma_1)\delta               \,,\\
    f &:= -(1-\gamma_1)\delta/(\alpha\beta) \,.
  \end{align}
\end{subequations}
Here we have  set
$\d=(\a\b-1)\g_1=(\a-\b)\g_2$.
The $a,\ldots,f$ satisfy the relation
$ab+cd=ef$. The matrix $\S_D$ reads
 \begin{align}
    \hat{S}_D
    &=
    \left(
    \begin{array}{rrrr}
      0 &  c &  e & -a \\
      c &  0 & -b &  f \\
      e &  b &  0 &  d \\
      a &  f &  d &  0 \\
    \end{array}
    \right)
    ,
  \end{align}
and has the characteristic polynomial,
\begin{align}
    \text{det}(\hat{S}_D-\lambda)
    &=
    (\lambda^2+a^2+b^2-c^2-d^2-e^2-f^2)~\,\lambda^2
    \,.  \nonumber
  \end{align}
Its kernel $\cal K$ is therefore two-dimensional.
With the relation $ab+cd=ef$ we obtain two linearly independent
basis vectors of $\cal K$,
 \begin{align}
    A^{(1)}&=\left(\begin{array}{r}  0 \\ -d \\  f \\  b \end{array}\right)
    ,&
    A^{(2)}&=\left(\begin{array}{r} -d \\  0 \\  a \\  e \end{array}\right)
     .\label{eq:eigenvects}
  \end{align}
The amplitudes in region $[120]$ follow from Eq.~(\ref{remamp}),
  \begin{align}
    \tilde{A}^{(1)}&=\left(\begin{array}{r}  -b \\ -f \\  d \\  0 \end{array}\right)
    ,&
    \tilde{A}^{(2)}&=\left(\begin{array}{r} -e \\  -a \\  0 \\  d \end{array}\right)
     .\label{remoteamps}
  \end{align}
This means  $\tilde{A}^{(j)}_{\s_1,\s_2}=-A^{(j)}_{-\s_1,-\s_2}$ and entails together with
the ansatz (\ref{BA}) the relation $f(x_1,x_2)=-f(-x_1,-x_2)$, i.e. the wave function
is of odd parity. Because the functions of the form given in
Eq.~(\ref{BA}) span the whole Hilbert space and are determined by a single region, the odd functions
corresponding to elements of $\cal K$ in region $[021]$ span the complete odd subspace. It follows that
the projection of the Hamiltonian onto the odd invariant subspace can be diagonalized with the Bethe ansatz, the partial waves show no diffraction and the system is therefore integrable if confined to this subspace.  
Let us note that the exceptional case
 $x_1=x_2=0$ has not to be considered separately as the wave function vanishes there and the Yang-Baxter equation 
for two-particle scattering is sufficient
to prove consistency of the ansatz (\ref{BA}).  

\begin{figure}[t b]
\includegraphics[width= 0.4\textwidth]{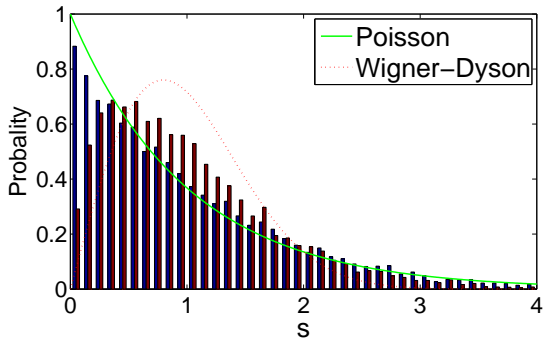}
\caption{(Color online) Level spacing statistics of the odd (blue bars) and even (red bars) subspace. The green solid line and the red dotted line indicate the Poisson $P_P(s)=\exp(-s)$ and Wigner-Dyson distribution for the orthogonal ensemble, $P_{WD}(s)= \frac{\pi s}{2} \exp(-\pi s^2/4)$, respectively. The parameters are $(V,U)=(2,2)$. The lattice has 201 sites with open boundary conditions. Each subspace contains roughly 10000 levels. 
\label{levelspacing}}
\end{figure}

The integrability of the odd subspace and the nonintegrability of the even subspace can be seen in Figs.~\ref{levelmol} and \ref{biclevel}, respectively, which depict parts of the spectral graph. There are apparent avoided crossings in the even subspace in Fig.~\ref{levelmol}, indicating nonintegrability according to the level-crossing criterion \cite{db}, while in Fig.~\ref{biclevel} the odd-parity bound state crosses both odd and even states. The latter is clear from parity symmetry, the former is due to integrability of the odd sector. 
This agreement between the numerical results on a finite lattice (which, however, preserves the parity symmetry) and our analysis of the infinite lattice hints that
the odd sector remains integrable on finite lattices. To look at this question from a different point of view, we show in Fig.~\ref{levelspacing} the statistics of the level spacing in each sector.  
The odd subspace appears to be
almost Poissonian, as predicted by 
the Berry-Tabor criterion for integrability \cite{ber-tab}. 
The even subspace, on the other hand, does not show Wigner-Dyson statistics---characteristic for
fully developed chaos \cite{chaos}. The statistics is close to Poissionian for large spacings as in the odd subspace, but deviates from both Poissonian and Wigner-Dyson for small spacings.  
Moreover, the numerical analysis reveals even states which have a close connection with
associated odd states (see Sec.~\ref{sectionaccom} below). 
The intermediate behavior of the level distribution in the 
nonintegrable sector may be related to 
a possible ``weakly diffractive'' dynamics which deserves further investigation.  
For example, it would be interesting to check whether the classical (continuum) limit 
of the model is ergodic or not. 

\subsection{Two Bethe form odd-parity bound states}

Having established the Bethe ansatz solvability of the model in the odd subspace, we now proceed to study the odd-parity localized states in the Bethe ansatz framework. Note that both the ground state and the molecule-type bound state encountered above are even and thus not in the Bethe form. 

To determine the conditions on $k_1$ and $k_2$ imposed by the normalizability of the wave function, it is convenient to
avoid explicit use of the symmetrizer $\cal S$ and go back to the ansatz in Eq.~(\ref{ansatz}).
For localized states, it is easy to see that among the four coefficients $B_{1\leq j \leq 4}$, only one can be finite. The reason is that, in region I$_2$, one of the two terms of $e^{ik_j x_j}$ and $e^{-ik_j x_j}$ diverges to infinity while the other converges to zero as $|x_j| \rightarrow \infty$. Therefore, their prefactors cannot be finite simultaneously. Without loss of generality, suppose $B_1=1$ while $B_2=B_3=B_4=0$. The (unnormalized) wave function in region I$_2$ then reduces to,
\begin{equation}\label{wave11}
f(x_1,x_2)= e^{ik_1x_1+ i k_2 x_2}- e^{-ik_2 x_1 -i k_1 x_2}.
\end{equation}
Since in region I$_2$, $x_1$ goes to $-\infty$ while $x_2$ goes to $+\infty$, we need $|e^{ik_1}|>1> |e^{ik_2}|$ for $f$ to vanish at infinity. Note that we can also define $-k_2$ as $k_1$, and $-k_1$ as $k_2$. To avoid counting a solution twice, we take the convention that $|e^{ik_1 +ik_2}|\leq 1$.
From Eq.~(\ref{oddb}), we get two independent equations for $U$, $V$, $k_{1,2}$,
\begin{subequations}\label{eqnuvk}
\begin{eqnarray}
\left(U+2i(\sin {k}_1-\sin {k}_2) \right) \left(1- \frac{V}{2i\sin {k}_1}\right)=0,\\
\left(U+2i(\sin {k}_1-\sin {k}_2) \right) \left(1+ \frac{V}{2i\sin{k}_2}\right)=0.
\end{eqnarray}
\end{subequations}
From (\ref{AB1}) and (\ref{AB2}), we get furthermore,
\begin{eqnarray}
A_1= 1-\frac{V}{2i \sin k_1}, \,  A_3=  \frac{V}{2i \sin k_1},\,A_2= A_4=0,\quad\quad \nonumber \\
A_5= A_6=0,\, A_7=  \frac{V}{2i \sin k_2},\, A_8=-1-\frac{V}{2i \sin k_2}.\quad \quad\label{a8}
\end{eqnarray}
In region I$_1$ ($0\leq x_1 \leq x_2$), $x_1$ and $x_{21}\equiv x_2-x_1$ are independent variables taking values on $\mathbb{Z}^+\cup 0$. Therefore, we rewrite the wave function as
\begin{eqnarray}
 &&f(x_1,x_2) =A_1 e^{i(k_1+k_2)x_1 +ik_2 x_{21}} + A_3 e^{i(k_2-k_1)x_1 + ik_2 x_{21}}\quad \nonumber\\
& &\quad \quad \, + A_7 e^{i(k_2-k_1)x_1 -ik_1 x_{21}} + A_8 e^{-i(k_1 +k_2)x_1 -ik_1 x_{21}}.\quad \label{wave3}
\end{eqnarray}
Now for the wave function to be localized, we need $|e^{ik_1 +ik_2}|<1 $ and $A_8=0$. 
We thus get from (\ref{a8})
\begin{equation}
-V=2i  \sin k_2 = e^{ik_2}- e^{-ik_2}.
\end{equation}
Eq.~(\ref{eqnuvk}b) is then satisfied. For (\ref{eqnuvk}a) to be satisfied, we need either $-V+2i \sin {k}_1=0 $ or $U+2i(\sin {k}_1-\sin {k}_2)=0$. But the former is unacceptable, otherwise the condition $|e^{ik_1}| >1$ and $|e^{ik_1 +ik_2}|<1$ can not be satisfied simultaneously. Therefore, $U+2i(\sin {k}_1-\sin {k}_2)=0 $, or
\begin{equation}\label{eff}
\quad -V-U=2i \sin {k}_1=e^{ik_1}-e^{-ik_1}.
\end{equation}
Defining $z_1=e^{ik_1}$ and $z_2=e^{ik_2}$, we have $
-V=z_2-z_2^{-1}$, $-V-U= z_1- z_1^{-1}$, and $E=-z_1-z_1^{-1}-z_2-z_2^{-1}$, with $ |z_2|<1<|z_1|<|z_2|^{-1} $. Instead of studying for what kind of $(V,U)$ pairs, we have solutions of $(z_1,z_2)$ satisfying this condition, we do the inverse. Because $V>0$ by assumption, we have $0<z_2<1$. Depending on the sign of $z_1$, we have two cases:

(i) $0< z_2<1<z_1<z_2^{-1}$. We get $ 0<-V- U = z_1-z_1^{-1} < z_2^{-1}- z_2 =V$. 
That is, $-2V < U < -V <0$. The energy of the wave function is
\begin{equation}\label{secondbs}
E_{b2}=- \sqrt{V^2 +4}- \sqrt{(V+U)^2+4}.
\end{equation}
It is easy to prove that $E_{b2}<\{-4,-\sqrt{V^2+4}-2,-\sqrt{U^2+16}\}$. Consequently, this bound state is below all the three continuum bands and is thus not a BIC. A notable feature of this state is that on the line $x_1+x_2=0$, $f(x_1,x_2)=0$ and if $x_1+x_2>0$, $f(x_1,x_2)<0$ while if $x_1+x_2<0$, $f(x_1,x_2)>0$. That is, the wave function has a node line $x_1+x_2=0$, and is positive on one of the two half-planes, while negative on the other. This property can be readily verified from the expressions \eqref{wave11} and \eqref{wave3}. 

\begin{figure}[tb]
\includegraphics[width= 0.35\textwidth, bb= 37   196   551   540]{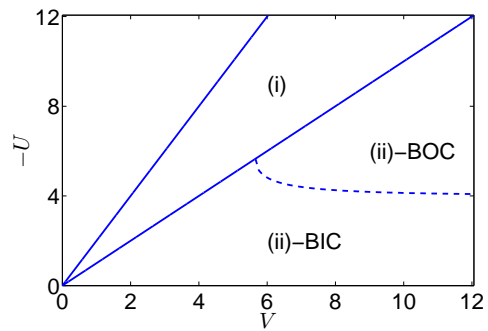}
\caption{(Color online) In the region labeled with (ii)-BIC, the state with energy $E_{b1}$ [see (\ref{bic})] is a bound state embedded in the $[-4,+4]$ continuum. In the region labeled with (ii)-BOC, this state is a bound state below the $[-4,+4]$ continuum band, and outside of all the three continuum bands. The boundary between the two regions is determined by the condition $E_{b1}=-4$. In the region labeled with (i), the odd-parity bound state with energy $E_{b2}$ exists, which is below all the continuum bands.
\label{bicboc}}
\end{figure}

(ii) $-z_2^{-1} <z_1 <-1<0<z_2<1$. We get
\begin{eqnarray}
0>-V- U = z_1-z_1^{-1} > -z_2^{-1}+ z_2 =-V. 
\end{eqnarray} 
That is, $-V < U <0$. The eigenenergy of the wave function is
\begin{equation}\label{bic}
E_{b1}=- \sqrt{V^2 +4}+ \sqrt{(V+U)^2+4}.
\end{equation}
Now, it is easy to show that $0>E_{b1}>\{ U, -\sqrt{V^2+4}+2 \}$, and thus $E_{b1}$ falls outside of the second and third bands. But \textit{it can fall in the continuum band $[-4,+4]$ to be an embedded eigenvalue.} For example, for $(V,U)=(2,-0.5)$, $E_{b1}=-0.3284$, which is inside the $[-4,+4]$ continuum. 
The condition for this state to be a BIC is $E_b>-4$. 
Note that this bound state exists whenever $V<U<0$. However, only in a subset [the region denoted as (ii)-BIC in Fig.~\ref{bicboc}] of this region does it fall in the continuum. Its energy can be tuned continuously across the edge $-4$ by tuning the parameters across the dashed line in Fig.~\ref{bicboc}. 

\begin{figure}[tb]
\centering
\includegraphics[ width= 0.4\textwidth]{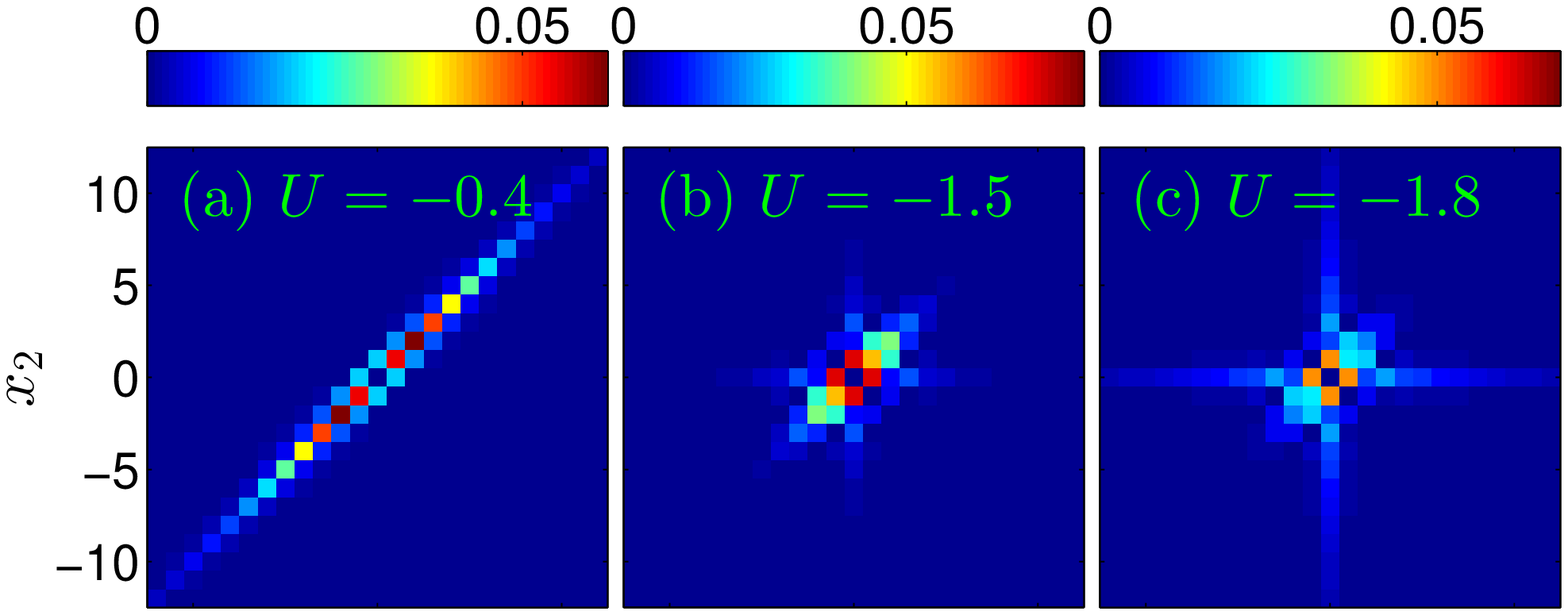}
\includegraphics[ width= 0.4\textwidth]{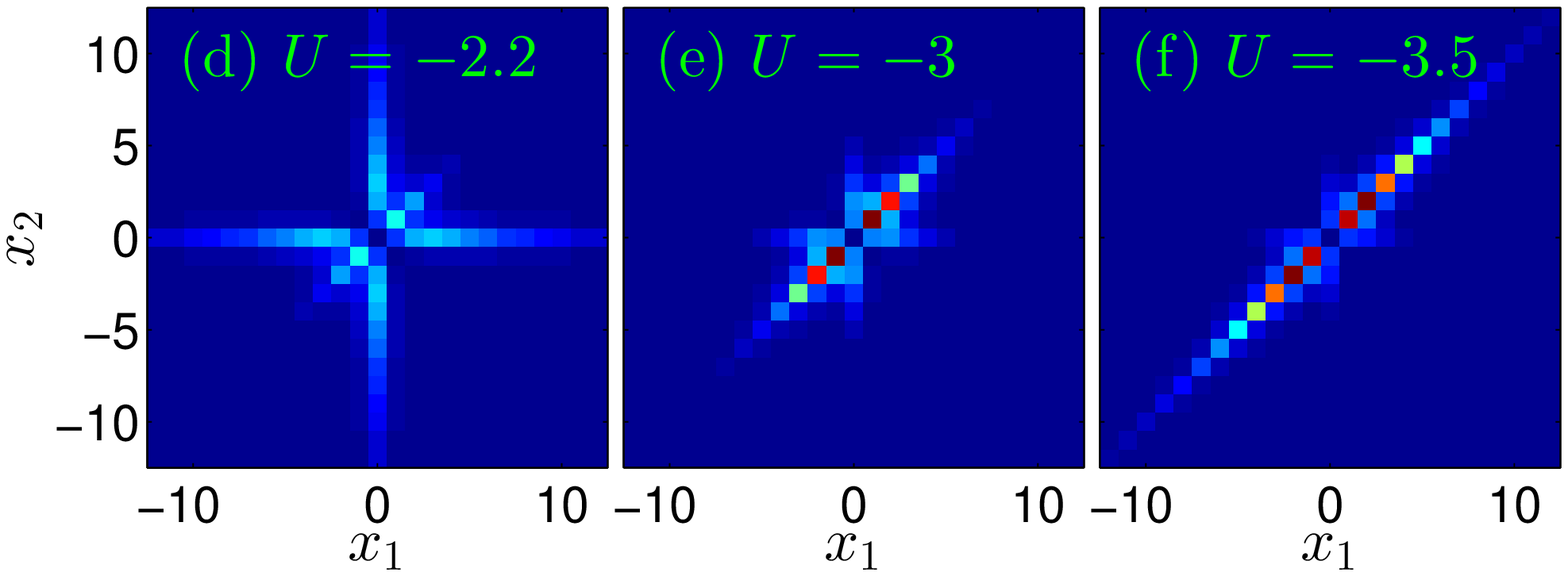}
\caption{(Color online) Images of the squared wave functions of the two analytically solvable bound states. The value of $U$ decreases monotonically from (a) to (f), with the value of $V$ fixed to $2$. Note that in (a)-(c), $-V<U<0$ while in (d)-(f), $-2V<U<-V$. These states are obtained in a finite lattice by exact diagonalization, with $100$ sites on each side of the impurity and open boundary conditions. They agree with the analytic expressions in Eqs.~(\ref{wave11}), (\ref{a8}), and (\ref{wave3}).
\label{bicwave}}
\end{figure}

In Fig.~\ref{bicwave}, we have plotted the squared wave functions of the two bound states with a fixed $V$ but different values of $U$. There we see that for a fixed value of $V$, the localized state becomes extended along the line of $x_1=x_2$ (see Fig.~\ref{bicwave}a) as $U\rightarrow 0^-$, while it becomes extended along the lines of $x_1=0$ and $x_2=0$ (see Fig.~\ref{bicwave}c) as $U\rightarrow -V^-$. This follows from $|z_1z_2|\rightarrow 1$ or $|z_1|\rightarrow 1$, respectively, in the two limits. Similar behavior is displayed by the other odd-parity bound state as $U\rightarrow -2 V^-$ or $ -V^+$, due to the same reason. As $U$ crosses $-V$ from $-V^-$ to $-V^+$ (see Figs.~\ref{bicboc}c and \ref{bicboc}d), the first bound state becomes extended and disappears, while the second one appears starting from an infinite size. However, the two states are not continuously linked. 
Actually, from (\ref{eff}) we see that $-V-U$ plays the role of an effective defect potential. As it changes sign, $z_1$ changes sign and the bound state jumps discontinuously from above the second band to below it. Finally, it should be stressed that in contrast to the power-law decay of the wave function in \cite{wigner}, here the wave function decays exponentially as $|x_{1,2}|$ tend to infinity, as seen in (\ref{wave11}) and (\ref{wave3}).  

\begin{figure}[tb]
\includegraphics[width= 0.4\textwidth]{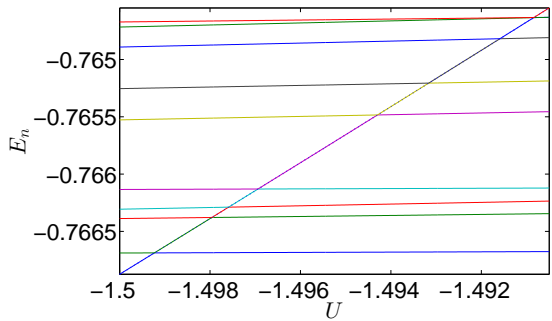}
\caption{(Color online) A small portion of the spectrum of the model with $ V=2 $. The lattice has 201 sites with open boundary conditions. In total, 10 levels are shown. The diagonal one corresponds to the bound state in the continuum (BIC). The horizontal lines correspond to extended states in the $[-4, +4]$ continuum band. Some of them are even, some are odd. The BIC level, which is odd, crosses both the even and odd ones. 
\label{biclevel}}
\end{figure}

Above we have shown that $E_{b1}$ falls in the continuum band for appropriate parameters. It would be instructive to see how this energy level moves inside the band as the parameters vary. Since the BIC is an eigenstate of the integrable sector, it is expected that $E_{b1}$ will cross the levels of the extended states without hybridization. This is indeed the case. In Fig.~\ref{biclevel}, we have shown a portion of the spectral graph of the model. We see that the BIC level (the diagonal one) is the only level that changes significantly in the interval of $U$ and it crosses all other (even or odd) levels leading to degeneracies at the crossing points. For even extended states
it is a consequence of the parity-symmetry, while for odd states it is expected from integrability of the odd sector \cite{db}.

\section{Two accompanying bound states}\label{sectionaccom}

\begin{figure}[tb]
\includegraphics[ width= 0.35\textwidth ]{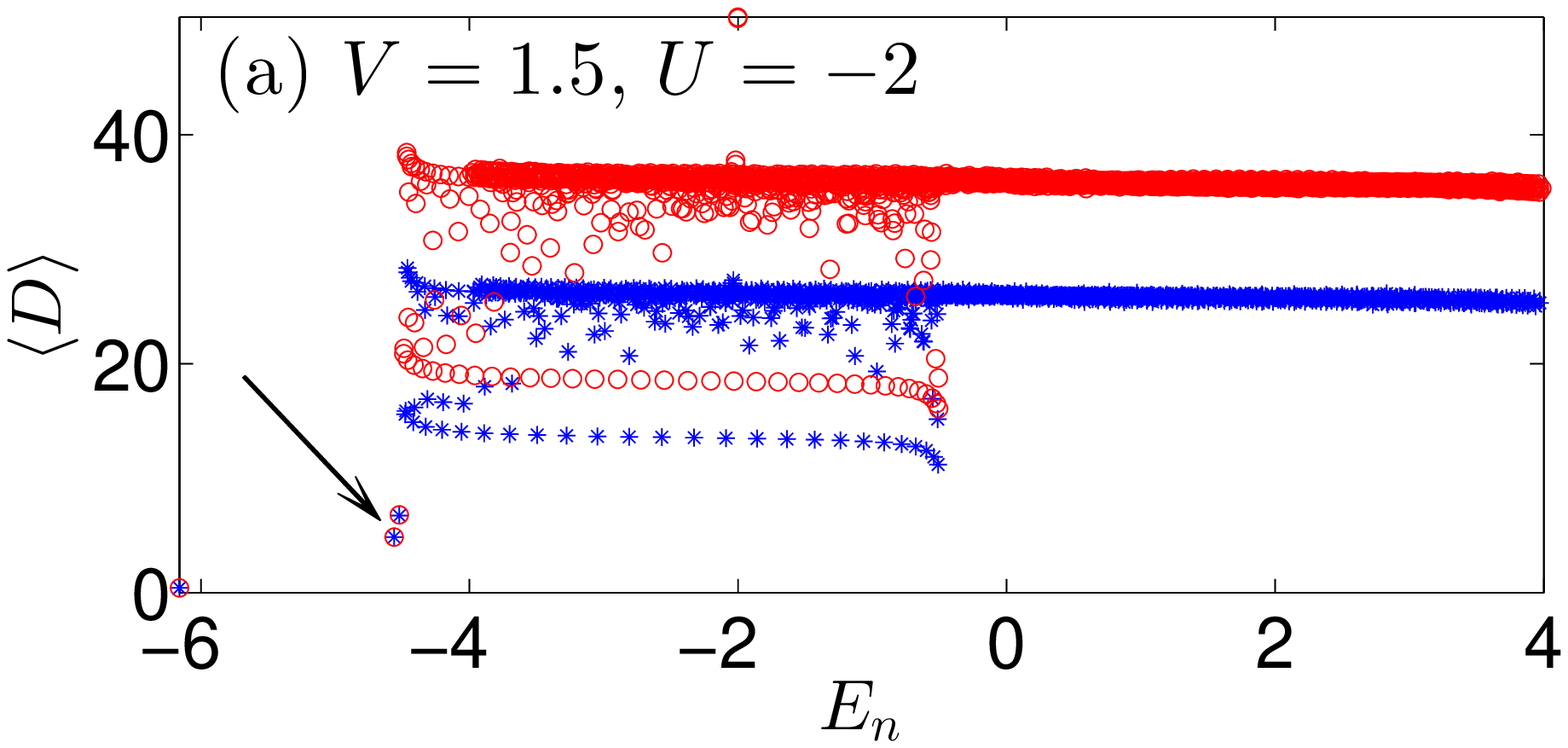}
\includegraphics[ width= 0.35\textwidth ]{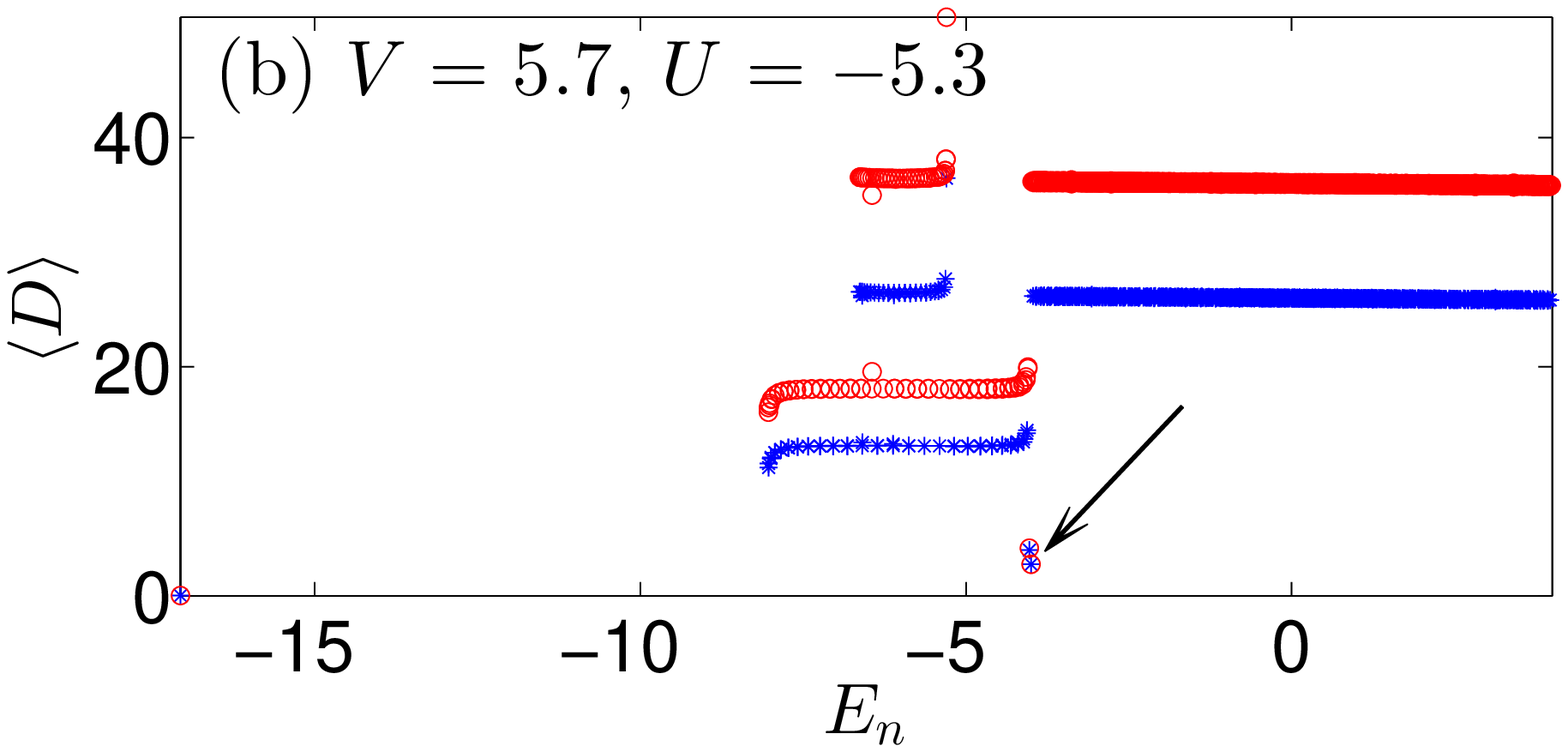}
\caption{(Color online) The two analytically solvable, odd-parity bound states and their accompanying even-parity bound states (each pair is indicated by an arrow) revealed by using the lattice expansion algorithm. The values of $(V,U)$ are shown in each panel. The upper panel corresponds to the case of $-2V<U<-V$, while the lower panel corresponds to the case of $-V<U<0 $. The smaller and larger lattices have  51 and 71 sites, respectively. \label{accom}}
\end{figure}

\begin{figure}[tb]
\centering
\includegraphics[ width= 0.4\textwidth]{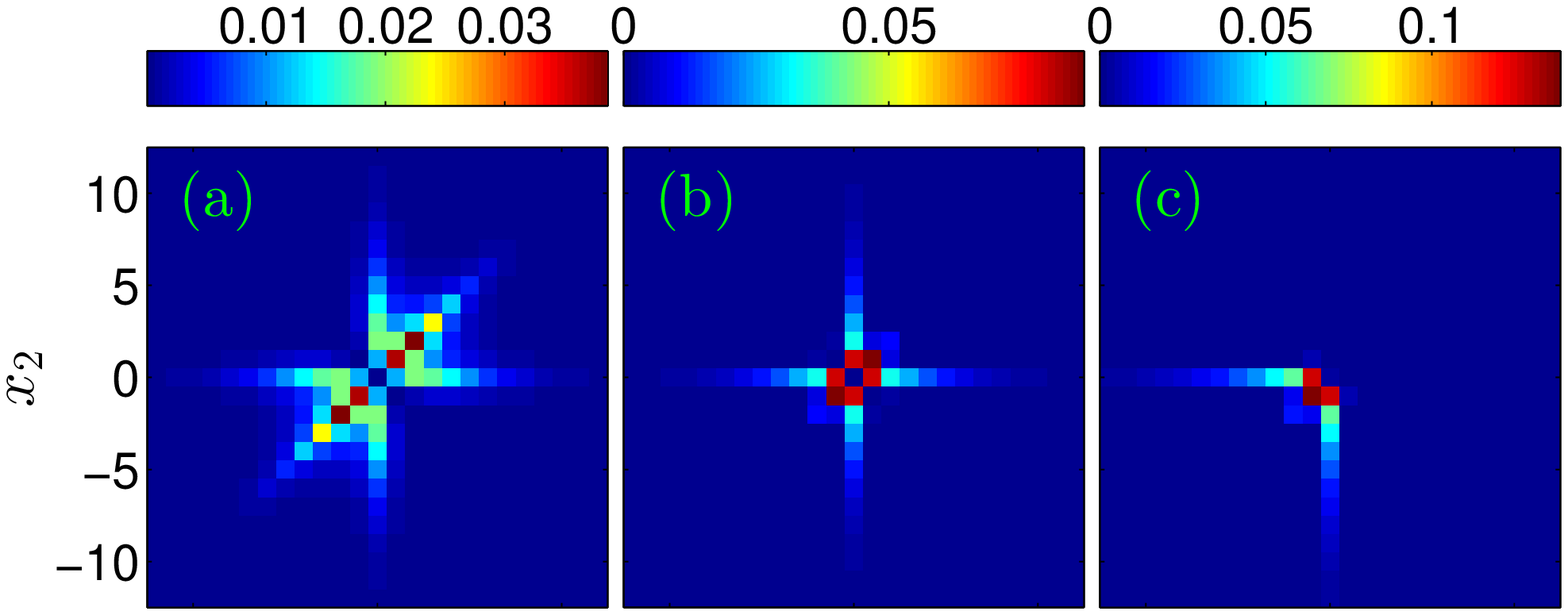}
\includegraphics[ width= 0.4\textwidth]{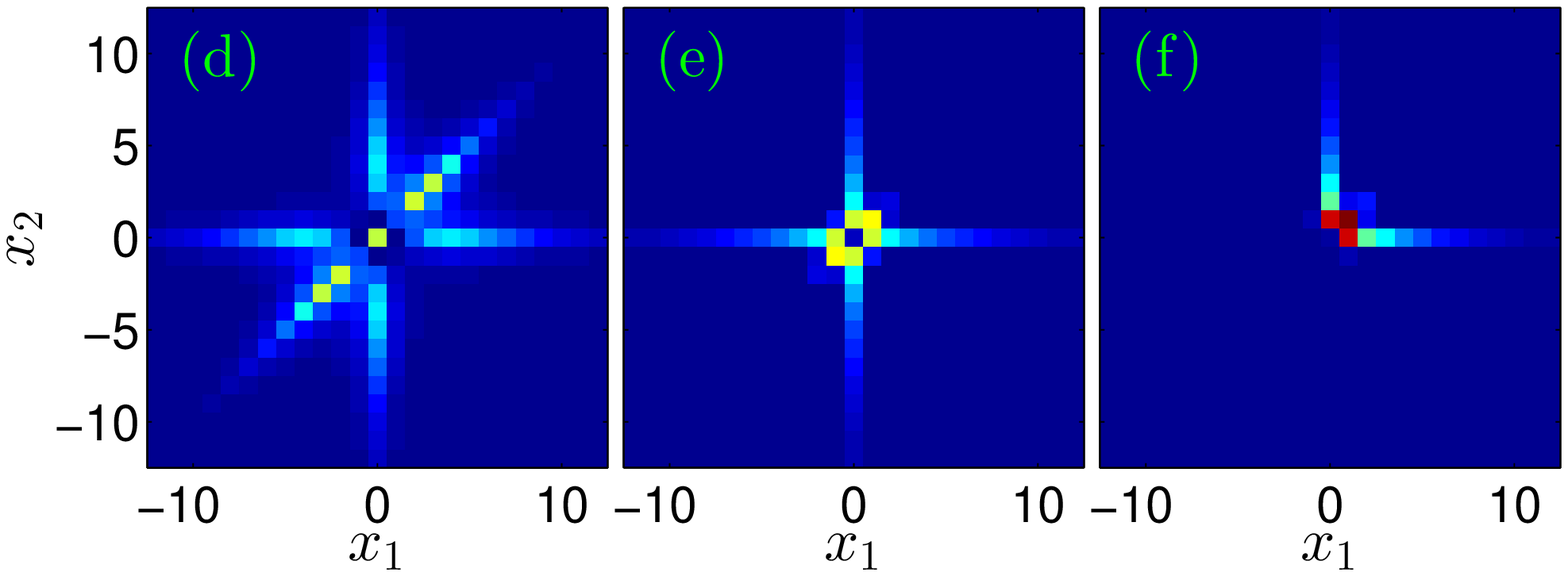}
\caption{(Color online) (a) and (b): Images of the squared wave functions of the two analytically solvable bound states in Fig.~\ref{accom}, and (d) and (e): their accompanying even-parity bound states. The values of $(V,U)$ are the same as in Fig.~\ref{accom}, i.e., $(V,U)=(1.5,-2)$ in (a) and (d), or $(V,U)=(5.7,-5.3)$ in (b) and (e). These states are obtained in a lattice with 201 sites and open boundary conditions by exact diagonalization. In (c) and (f), the squared wave function of the sum or the difference of the two bound states in (b) and (e) are shown, respectively. 
\label{twoaccomwave}}
\end{figure}

In the proceeding section, we have solved two odd-parity bound states analytically, each of which appears in an appropriate region. They appear in Figs.~\ref{fig3}b and \ref{fig3}c. Numerically, it is observed that often each of the two states is accompanied by an even-parity bound state, which is close to it in energy. Actually, in Fig.~\ref{fig3}b, one of the excited bound states has odd-parity, while the other has even-parity. In Fig.~\ref{fig3}c, we see only the odd-parity one (the BIC). However, if its energy is tuned outside the $[-4,+4]$ continuum band, it can have an accompanying even-parity bound state. In Fig.~\ref{accom}, we show two cases when the two odd-parity bound states each has an accompanying even-parity bound state. 

Unlike the odd-parity bound states, the even-parity bound states have not the simple Bethe form.
This makes them much more difficult to study and understand. However, thorough numerical exploration reveals that they are closely related to their odd-parity companions. First, in each pair, the two states are ordered in a definite way and the gap between them is always much smaller than unity. Specifically, the odd-parity state with energy $E_{b2}$ (if it exists) is always the first excited state, while its companion is (if it exists) the second excited one. 
For the odd-parity state with energy $E_{b1}$, its companion (if it exists) is always lower in energy to it. Second, as illustrated in Fig.~\ref{twoaccomwave}, the wave functions of the two bound states in each pair are similar to each other in profile. More precisely, if we denote the wave functions of the even and odd states in each pair as $f_{e}$ and $f_o$, respectively, then $f_e+f_o$ and $f_e-f_o$ are well localized on one of the two half-planes $x_1+x_2 \lessgtr 0$. This is demonstrated in Figs.~\ref{twoaccomwave}c and \ref{twoaccomwave}f. These two facts remind us of the double-well potential, where we have pairs of even and odd eigenstates, which are nearly degenerate in energy and can be expressed as linear combinations of left- and right- localized modes. It indicates that the wave function of the even-parity bound state, while not exactly in the Bethe form, may be nevertheless close to it, a numerical finding which should be further studied.  

\begin{figure}[tb]
\includegraphics[width= 0.35\textwidth]{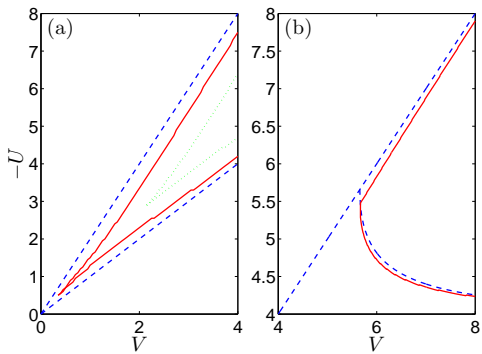}
\caption{(Color online) (a) In the region enclosed by the red lines, an accompanying even-parity bound state to the analytically solvable odd-parity bound state with energy $E_{b2}$ exists. The green dotted lines enclose the region where the simple variational approach predicts the existence of such an accompanying bound state. The blue dashed lines enclose the region $-2V<U<-V$ where the $E_{b2}$ bound state exists. (b) In the region enclosed by the red lines, an accompanying even-parity bound state to the analytically solvable odd-parity bound state with energy $E_{b1}$ exists. The $E_{b1}$ bound state exists below the straight blue dashed line and moves into the continuum below the curved line (comp. Fig.~\ref{bicboc}).
\label{evenregion}}
\end{figure}

We have tried to map out the regions where the two accompanying bound states appear. The results are shown in Fig.~\ref{evenregion}. In Fig.~\ref{evenregion}a, we see that if $U$ is too close to the edges of the region $-2V<U<-V$, the even-parity bound state disappears. The reason is that in these two limits, $E_{b2}$ gets close to the lower edges of the second or third band [see Eq.~\eqref{secondbs}]. The even-parity one would get even closer or just inside the bands and thus becomes unstable. As for the second even-parity bound state, its appearance region is approximately the region where the $E_{b1} $ odd-parity bound state is a BOC. As $E_{b1}$ gets too close to the upper edge of the second band ($U \rightarrow -V^-$), or too deep inside the $[-4 ,+4]$ band, the even-parity companion becomes unstable and disappears.

Here, we would like to remark that for the odd-parity bound state with energy $E_{b2}$, the presence of an accompanying even-parity bound state is expected in some region. As mentioned in Sec.~\ref{twoEX}B, the wave function of the odd-parity bound state $f_{b2}$ vanishes on the line $x_1+x_2=0$ and is completely positive on one of the planes $x_1+x_2 \lessgtr 0$ while completely negative on the other. Now construct an even-parity state
\begin{eqnarray}
f_{tr}= \begin{cases}  +f_{b2}(x_1,x_2) , & x_1+x_2\geq 0 ,\\ -f_{b2}(x_1,x_2) , & x_1+ x_2<0 .\end{cases}
\end{eqnarray}  
By Eq.~(\ref{h22}), we have then $\langle f_{tr}| \hat{H} | f_{tr}\rangle =\langle f_{b2}| \hat{H} |f_{b2}\rangle =E_{b2}$, which is lower than the lower edges of all the three bands. Then construct another even-parity state $\tilde{f}_{tr}=(1-|g\rangle \langle g|)f_{tr}$. This state is even and orthogonal to both the ground state $|g\rangle$ and the odd-parity state $f_{b2}$. If its energy is lower than the lower edges of the three continuum bands, then there must exist another even-parity bound state. This is likely since if $-2V<U<-V<0$, the ground state $|g\rangle $ localizes sharply at the origin where $f_{tr}$ is zero and thus $\tilde{f}_{tr}$ is actually very close to $f_{tr}$.
According to this criterion, we can prove that in the region enclosed by the green dotted lines in Fig.~\ref{evenregion}a, there should be an even-parity bound state. 

\section{Conclusions and Outlook}\label{sectionconclusion}

We have investigated the bound states in a system of two interacting bosons in a  one-dimensional lattice with a defect. 
The aim was to study a problem known from atomic physics  on a lattice. 
As the simplest one among its type, our model is both analytically and numerically tractable. 
Somewhat unexpectedly, this simple  model reveals rich physics.

First, we find that the repulsion between the two bosons plays a dual role. On one hand, as intuitively expected, a sufficiently large $U$ can delocalize the ground state. On the other hand, a sufficiently large $U$ 
can create a bound state and stabilize it \cite{reentrant}. Quantitatively, there exist two critical values $U_{c1,2}$ ($0<U_{c1}<U_{c2}$) of $U$. For $U>U_{c1}$, the ground state is no longer localized. However, for  $U>U_{c2}$, a molecule-type bound state appears and becomes  more and more  localized as $U$ increases. The large $U$ can in fact bind the two bosons into a molecule, which hops on the lattice with an effective amplitude $\propto 1/U$, and can thus be trapped by the defect. This localization becomes sharper with increasing $U$ because the effective hopping
decreases as $1/U$. 

An interesting point is that the two transitions have distinct natures. For the former, the ground state energy and wave function merge into the continuum continuously as $U$ increases towards $U_{c1}$. Near $U_{c1}$, the wave function can be well approximated by a modified Chandrasekhar wave function and the value of $U_{c1}$ can be accurately predicted in this way. In contrast, for the latter, the bound state energy merges into the continuum by a sharp anti-crossing as $U$ decreases towards $U_{c2}$, which means that the bound state becomes extended almost abruptly. This bound state is a \textit{bound state at threshold} \cite{quantum,zeroeig,volcano,mattisrmp,hoffmann}, which has a finite size even at the threshold of the continuum. 

Second, we have gained a better understanding of the analytical structure of the model. Though the model was believed to be nonintegrable \cite{mcguire}, we have seen that the odd-parity subspace is integrable. This makes it possible to write down the wave functions, in particular, the two bound states, explicitly in the Bethe form. Remarkably, one of the two turns out to be a \textit{bound state embedded in the continuum} (BIC). It is the simplest BIC we know so far, in terms of the energy and wave function; and most importantly, it arises in a non-engineered model. Therefore, in this simple model, there can be two types of exotic bound states: one remains finite at the threshold of the continuum; the other resides inside the continuum.

Eigenstates in the even-parity subspace do not have the Bethe form. However, the fact that the level spacing statistics in this subspace interpolates between the Poisson distribution and the Wigner-Dyson distribution, 
hints at a \textit{weakly diffractive} dynamics even in this non-integrable sector, which should be investigated further
in the future.

Third, besides the four types of bound states mentioned above, we have found numerically that in some regions of the parameter space the  odd-parity bound states may possess accompanying, even-parity bound states, i.e. there are two bound states of different parity which are close in energy and have similar wave functions. 
The situation resembles the nearly degenerate pairs of eigenstates in a double-well potential. Again, this phenomenon implies that there exists some relation yet to be discovered between the diffractive even sector and the non-diffractive odd sector. 

In this paper, we have focused on the properties of the bound states, which is mainly a static problem. In the future, it is worthwhile to study the dynamics of the model \cite{weiss,kolovsky}. In fact, the analogy of the model with some atomic physics problems can be generalized to the dynamics. For example, consider the following: Initially one boson is trapped in the defect mode and the other boson comes in from faraway. What would the system evolve into? There are many possibilities. First, the injecting boson can be reflected or transmitted without exciting the initially trapped boson. Second, the initially trapped boson may get excited and both bosons become free and move independently. Third, the two bosons can go away as a molecule. It is a nontrivial task to find the probability for each of these possibilities. The problem obviously bears analogy to the well-known problem of electron scattering off hydrogen atoms \cite{callaway}.  

\section*{Acknowledgment}

We are grateful to Jiangping Hu for his interest in this problem and stimulating discussions, Michael Sekania and Dieter Vollhardt for their helpful comments. This work was supported in part by the Deutsche Forschungsgemeinschaft through TRR 80.

\appendix

\section{Variational proof of $U_{c1}\geq V$}\label{variational}

We argue that if $U<V$, the ground state is a bound state. To this end, we just need to prove that the ground state energy is below all the three continuum bands, since any eigenvalue outside of the continuum bands should be a bound state according to our argument in Sec.~\ref{thebhm}.

Consider a direct product variational wave function $f(x_1,x_2)=\phi_1(x_1)\phi_2(x_2)$, with $\phi_{1,2}$ normalized to unity. Its energy $E\equiv \langle f| \hat{H} | f\rangle$ is
\begin{eqnarray}
E  = \langle \phi_1|h|\phi_1\rangle + \langle \phi_2|h|\phi_2\rangle  +U\sum_{x} \phi_1^2(x)\phi_2^2(x).\quad 
\end{eqnarray}
Here $h$ is the single particle Hamiltonian, i.e., the sum of the hopping and the potential part of the full Hamiltonian. Now it is clear that particle 1 sees an effective potential $
W_1(x)=-V\delta_{x,0}+ U \phi_2^2(x)$, which depends on the wave function of particle 2, while particle 2 sees an effective potential $W_2(x)=-V\delta_{x,0}+ U \phi_1^2(x)$, which in turn depends on the wave function of particle 1. Regardless of the concrete forms of $\phi_{1,2}$, we have always $\sum_m W_1(m)=\sum_{m}W_2(m)=U-V$.
Variationally we know that if $U<V$, then both $W_1$ and $W_2$ admit localized ground states. Now choose $\phi_1$ as the ground state of $h$ with eigenvalue $-\sqrt{V^2+4}$, and then choose $\phi_2$ as the ground state, a localized state, in the effective potential $W_2$. We have then,
\begin{equation}\label{e}
E=\langle \phi_1|h|\phi_1\rangle + \langle \phi_2|h+W_2|\phi_2\rangle < -\sqrt{V^2+4}-2.
\end{equation}

We can decompose the direct product wave function as the superposition of a symmetric (bosonic) and an antisymmetric (fermionic) function, $ f(x_1,x_2) = f_{s}(x_1,x_2)+ f_{a}(x_1,x_2) $, with
\begin{eqnarray}
f_s(x_1,x_2)= \frac{1}{2} \left(\phi_1(x_1)\phi_2(x_2)+\phi_2(x_1)\phi_1(x_2) \right), \label{fs}\\
f_a(x_1,x_2)= \frac{1}{2} \left(\phi_1(x_1)\phi_2(x_2)-\phi_2(x_1)\phi_1(x_2)\right).
\end{eqnarray}
Note that $f_{s}$ and $f_{a}$ are not normalized. Actually, $
1=\langle f|f\rangle =\langle f_{s}|f_{s}\rangle + \langle f_{a}|f_{a}\rangle
$. The even and odd functions do not mix under the action of $\hat{H}$. Therefore,
\begin{equation}
\langle f|\hat{H} |f \rangle =\langle f_{s}|f_{s}\rangle \frac{\langle f_{s}|\hat{H}| f_{s}\rangle}{\langle f_{s}| f_{s}\rangle} + \langle f_{a}| f_{a}\rangle\frac{\langle f_{a}|\hat{H} | f_{a}\rangle}{\langle f_{a}| f_{a}\rangle}.
\end{equation}
We note that for a fermionic wave function as $f_{a}$, the interaction is non-effective. Therefore, by Pauli principle, we have 
\begin{equation}\label{odde}
\frac{\langle f_{a}|\hat{H} | f_{a}\rangle}{\langle f_{a}| f_{a}\rangle} \geq -\sqrt{V^2 +4}-2.
\end{equation}
From (\ref{e}) and (\ref{odde}), we have 
\begin{equation}\label{evene}
\frac{\langle f_{s}|\hat{H} | f_{s}\rangle}{\langle f_{s}| f_{s}\rangle} < -\sqrt{V^2 +4}-2.
\end{equation}
Therefore, we have constructed a localized bosonic state $f_{s}$, which is below all the three bands. Thus, the ground state is a localized state if $U<V$. We have thus obtained a lower bound for $U_{c1}$, i.e., $U_{c1} \geq  V$.

\section{ Lower limit of $U_{c1}$ by Chandrasekhar trial wave functions}\label{chandrasekhar}
We employ the Chandrasekhar trial wave function
\begin{equation}
f(x_1,x_2)=e^{-\alpha_1 |x_1| -\alpha_2 |x_2|}+e^{-\alpha_2 |x_1| -\alpha_1 |x_2|},
\end{equation}
with two variational parameters satisfying $0\leq \alpha_1\leq \alpha_{2}$. Note that the wave function is in the form of $f_s$ [see Eq.~\eqref{fs}] in App.~\ref{variational}, but with the orbits $\phi_{1,2}$ specified.

We have for the norm of the wave function
\begin{eqnarray}
\langle f|f \rangle
&=& \frac{2}{\tanh \alpha_1 \tanh \alpha_2} + \frac{2}{\tanh \bar{\alpha} \tanh \bar{\alpha}},
\end{eqnarray}
with $\bar{\alpha}=(\alpha_1 +\alpha_2)/2$ and $\delta = (\alpha_2 -\alpha_1)/2$.
We have also
\begin{eqnarray}
\langle f|\hat{H} |f \rangle &=& - \frac{4}{\sinh \alpha_1 \tanh \alpha_2} - \frac{4}{ \sinh \alpha_2 \tanh \alpha_1}  \nonumber \\
&& - \frac{8 \cosh \delta}{\sinh \bar{\alpha} \tanh \bar{\alpha}} +\frac{4U}{\tanh (\alpha_1 +\alpha_2)} \nonumber\\
&& - 2V \left( \frac{1}{\tanh \alpha_1}+ \frac{1}{\tanh \alpha_2}+ \frac{2}{\tanh \bar{\alpha}} \right) .\quad 
\end{eqnarray}
The energy of the trial wave function is then $
E=\langle f | \hat{H} |f \rangle/\langle f|f \rangle= (a+b)/(c+d)$,
with 
\begin{eqnarray}
a &=& - \frac{4}{\sinh \alpha_1 \tanh \alpha_2} - \frac{4}{ \sinh \alpha_2 \tanh \alpha_1}- \frac{2V}{\tanh \alpha_1} , \quad \,\\
b&=& - \frac{8 \cosh \delta}{\sinh \bar{\alpha} \tanh \bar{\alpha}} +\frac{4U}{\tanh (\alpha_1 +\alpha_2)} \nonumber \\
&& - 2V \left(  \frac{1}{\tanh \alpha_2}+ \frac{2}{\tanh \bar{\alpha}} \right) \\
c&=& \frac{2}{\tanh \alpha_1 \tanh \alpha_2}, \quad 
d= \frac{2}{\tanh \bar{\alpha} \tanh \bar{\alpha}}.
\end{eqnarray}
Since $c,d>0$, we have $E\geq \min(a/c,b/d)$. However, it is easy to prove that $a/c\geq E_{edge2}$ by noting that $\sinh(\alpha_1)\geq \tanh(\alpha_1)$. Therefore, we need $b/d\leq E_{edge2}$ to make $E\leq E_{edge2}$ possible. Numerical evidence indicates that in this variational framework, at the critical point of $U=U_{c1}$, $\alpha_1=0$ and $\alpha_2=\ln [\frac{1}{2}(V+\sqrt{V^2+4})]$. 
That is, one boson is delocalized while the other resides in the single-particle localized mode corresponding to the defect potential. 

The critical value $U_{c1}$ is then determined by the condition $b/d=E_{edge2}$ with $\alpha_{1,2}$ taking the values above. It is straightforward to obtain the limiting behaviors of $U_{c1}$:
\begin{eqnarray}
U_{c1}= \begin{cases}  \frac{3}{2}V, & V\rightarrow 0 ,\\ V+1, & V \rightarrow +\infty ,\end{cases}
\end{eqnarray}
It is an improvement over the variational approximation in App.~\ref{variational} in both limits. However, it is still inferior to the modified Chandrasekhar variational approximation.


\begin{thebibliography}{99}
\bibitem{definition}
It should be stressed that in this paper, by bound states or localized states we mean in fact normalizable states. They are not to be confused with extended molecule states in which the atoms are bound with respect to each other but the center-of-mass moves extendedly.
\bibitem{ipr}
Of course, there exist many other quantities, such as the inverse participation ratio, which can serve the same purpose.
\bibitem{forbidden}
Coupling to the other state $\frac{1}{\sqrt{2}}(|0,1\rangle- |-1,0\rangle )$ is forbidden by symmetry.
\end{thebibliography}
\end{document}